\documentclass[prb,twocolumn,floatfix]{revtex4}

\usepackage{amssymb, amsmath}
\usepackage{mathrsfs}
\usepackage[usenames]{color}
\usepackage{simplewick}
\usepackage{subfigure, floatflt}
\usepackage[dvips]{graphicx}

\usepackage[dvips]{hyperref}

\newcommand{\nquad}{\!\!\!\!\!}

\newcommand{\func}[2]{\operatorname{#1}(#2)}

\newcommand{\ket}[1]{| #1 \rangle}
\newcommand{\bra}[1]{\langle #1 |}

\newcommand{\avg}[1]{\langle #1 \rangle}
\newcommand{\op}[1]{\hat{#1}}
\newcommand{\opdag}[1]{\hat{#1}^\dagger}
\newcommand{\up}{\uparrow}
\newcommand{\dn}{\downarrow}

\newcommand{\reference}[5]{#1, #2 {\bf #3}, #4 (#5)}
\newcommand{\RMP}{Rev. Mod. Phys.}

\newcommand{\PRB}{Phys. Rev. B}
\newcommand{\PRL}{Phys. Rev. Lett.}
\newcommand{\PR}{Phys. Rev.}
\newcommand{\SST}{Supercond. Sci. Tech.}
\newcommand{\PTP}{Prog. Theor. Phys.}
\newcommand{\JPSJ}{J. Phys. Soc. Jp.}
\newcommand{\AnnP}{Annals of Phys.}
\newcommand{\AdvP}{Adv. in Phys.}
\newcommand{\ZPB}{Z. Phys. B}
\newcommand{\etal}{\textit{et.\@ al.\/}}

\newcommand{\Hn}{\hat{\mathcal{H}}}

\newcommand{\gutz}{\hat{P}}

\newcommand{\vv}[1]{\mathbf{#1}}
\newcommand{\nn}[1]{\bar{#1}}
\newcommand{\set}[1]{\mathcal{#1}}

\begin{document}

\title{Extended Gutzwiller Approximation for Inhomogeneous System}
\author{Wing-Ho Ko}
\author{Cody P. Nave}
\altaffiliation{Current Address: Condensed Matter Theory Center, Department of Physics, University of Maryland, College Park, Maryland 20742-4111, USA }
\author{Patrick A. Lee}
\affiliation{Department of Physics, Massachusetts Institute of Technology, Cambridge,
Massachusetts 02139, USA }
\date{December 11, 2007}

\begin{abstract}
The generalization of the Gutzwiller approximation to inhomogeneous systems is considered, with extra spin-and-site-dependent fugacity factors included. It is found that the inclusion of fugacity factors reconciles the seemingly contradictory choices of Gutzwiller factors used in the literature. Moreover, from the derivation of the Gutzwiller factors, it is shown that the Gutzwiller approximation breaks the rotational symmetry of the trial wavefunctions, and that different components of the spin-spin interaction need to be renormalized differently under the approximation. Various schemes to restore the rotational symmetry are discussed and are compared with results from variational Monte-Carlo calculations for the two-dimensional square-lattice antiferromagnet. Results along different paths within the full parameter space, which corresponds to different choices of fugacity factors in the literature, are also compared.
\end{abstract}

\maketitle

\section{Introduction}
The $t$-$J$ model in two dimensions has long been a focus in condensed matter physics, for despite its seemingly simple appearance, it is believed to describe a variety of important strongly-correlated electronic systems, including the quantum antiferromagnets, the spin liquids, and the high-temperature superconductors.\cite{SpinLiquid} The $t$-$J$ model is described by the Hamiltonian:
\begin{equation} \label{eq: Hamiltonian}
\Hn = \sum_{i,j,\sigma} t_{ij} \, \gutz \op{c}^\dagger_{i\sigma} \op{c}_{j\sigma} \gutz
+ \sum_{\avg{ij},\sigma} J \, \op{\vv{S}}_i \cdot \op{\vv{S}}_j
\end{equation}
where $\gutz = \prod_j (1-\op{n}_{j\up} \op{n}_{j\dn})$ is the \emph{Gutzwiller projection operator}, which accounts for the physical on-site Coulomb repulsion by explicitly prohibiting double occupation on any site.

A common way to study the $t$-$J$ model is to employ a variational approach, by introducing the \emph{Gutzwiller trial wavefunction}:
\begin{equation}
\ket{\psi} = \gutz \ket{\psi_0}
\end{equation}
where $\ket{\psi_0}$ is in general a Hartree-Fock type wavefunction that need not respect the double occupation constraint.

With the Hamiltonian $\Hn$ and the trial wavefunction $\ket{\psi}$, various expectations of the system can in principle be calculated numerically by the variational Monte Carlo (VMC) method.\cite{Gros} However, the VMC method is computationally costly and inefficient when long-range order of the system is sought, where the number of parameters in the trial wavefunction $\ket{\psi}$ becomes large.

A more practical way to perform these calculations is to employ the \emph{Gutzwiller approximation},\cite{EdeggerReview} first introduced by Gutzwiller,\cite{Gutz} and subsequently clarified and extended by Ogawa \etal,\cite{Ogawa} Vollhardt,\cite{Vollhardt} and Zhang \etal\cite{Zhang} In the approximation, an expectation with respect to  $\ket{\psi}$ is approximated by multiplying the expectation with respect to $\ket{\psi_0}$ by a factor that accounts for double occupation exclusion. i.e.,\@
\begin{equation} \begin{aligned}
\avg{\opdag{c}_{i\sigma} \op{c}_{j\sigma}} & \approx g_t(i,j,\sigma) 
	\avg{\opdag{c}_{i\sigma} \op{c}_{j\sigma}}_0 \\
\avg{\op{\vv{S}}_i \cdot \op{\vv{S}}_j} & \approx g_J(i,j) \avg{\op{\vv{S}}_i \cdot \op{\vv{S}}_j}_0
\end{aligned} \end{equation}
where $\avg{\op{Q}} = \bra{\psi}\op{Q}\ket{\psi} / \avg{\psi | \psi}$ and $\avg{\op{Q}}_0 = \bra{\psi_0}\op{Q}\ket{\psi_0} / \avg{\psi_0 | \psi_0}$ for any operator $\op{Q}$. The expectation $\avg{\op{Q}}$ of the \emph{projected} wavefunction $\ket{\psi}$ is said to be \emph{renormalized} from the expectation $\avg{\op{Q}}_0$ of the \emph{pre-projected} wavefunction $\ket{\psi_0}$.\cite{Zhang}

Generally, the \emph{Gutzwiller factors} $g_t(i,j)$ and $g_J(i,j)$ are obtained by ignoring any non-combinatorial configuration dependences of the expectation values. In the literature, for a homogeneous paramagnetic system, i.e., when $\avg{\op{n}_{j\up}}_0 = \avg{\op{n}_{i\up}}_0 = \avg{\op{n}_{j\dn}}_0 = \avg{\op{n}_{i\dn}}_0$ for all sites $i$, $j$, the values of $g_t$ and $g_J$ are known to be:\cite{Zhang,Edegger,Fukushima}
\begin{equation}\label{eq: Gutz-homo}
\begin{aligned}
g_t & = 2\delta / (1+\delta)\\
g_J & = 4/(1+\delta)^2
\end{aligned}\end{equation}
where $\delta = 1-\avg{\op{n}_{i\dn}}_0 - \avg{\op{n}_{i\up}}_0$. Alternatively, Eq.~\ref{eq: Gutz-homo} can also be derived from a functional integral approach.\cite{Kotliar}

However, the generalization of the two factors in an inhomogeneous case is not obvious. Some authors\cite{Huang,Poilblanc,Ziqiang} simply take the results from Eq.~\ref{eq: Gutz-homo} and interpret $\delta$ as site dependent. Hence, they obtain:
\begin{equation}\label{eq: Gutz-adhoc}
\begin{aligned}
g_t(i,j) & = \sqrt{4\delta_i \delta_j/(1+\delta_i)(1+\delta_j)}\\
g_J(i,j) & = 4/(1+\delta_i)(1+\delta_j)
\end{aligned}\end{equation}
where $\delta_i = 1-\avg{\op{n}_i}_0$ is the local hole density at site-$i$. 

In contrast, for an anti-ferromagnetic trial wavefunction in a square lattice, Gan \etal\cite{Gan} obtained:
\begin{equation}\label{eq: Gutz-y=1}
\begin{aligned}
g_t(A,B) & = n(1-n)/(n-2 n_{+} n_{-})\\
g_J(A,B) & = n^2/(n-2 n_{+} n_{-})^2
\end{aligned}\end{equation}
where $A$, $B$ are labels for sublattices, and which $n_{+} = \avg{\op{n}_{A\up}}_0 = \avg{\op{n}_{B\dn}}_0$, $n_{-} = \avg{\op{n}_{A\dn}}_0 = \avg{\op{n}_{B\up}}_0$, $n = n_{+} + n_{-}$. It should be noted that the $g_t$ they obtained is identical to that of Ogawa \etal,\cite{Ogawa} who did not derive $g_J$.

Another suggestion for $g_t$ and $g_J$ is given by Wang \etal,\cite{Wang} who derived their results from grand-canonical wavefunctions and considered a generalized Gutzwiller projection operator $\gutz' = \prod_j y_j^{\op{n}_j} (1-\op{n}_{j\up} \op{n}_{j\dn})$, where $y_j$ are local fugacities determined by the condition that $\avg{\op{n_j}}_0 = \avg{\op{n_j}}$. For the $t$-$J$ model, the factors they obtained are of the form $g_t(i,j,\sigma) = \sqrt{g_t}(i,\sigma) \sqrt{g_t}(j,\sigma)$ and $g_J(i,j) = \sqrt{g_J}(i) \sqrt{g_J}(j)$, where:
\begin{equation}\begin{aligned} \label{eq: Gutz-Wang}
\sqrt{g_t}(i,\sigma) & = \sqrt{\frac{(1-n_{i\nn{\sigma}}) (1-n_{i\up}-n_{i\dn})(n_{i\up}+n_{i\dn})}{(1-n_{i\sigma})(n_{i\up}+n_{i\dn}-2 n_{i\up}n_{i\dn})}} \\
\sqrt{g_J}(i) & = \frac{n_{i\up}+n_{i\dn}}{n_{i\up}+n_{i\dn}-2 n_{i\up}n_{i\dn}}
\end{aligned}\end{equation}
with $n_{i\sigma} = \avg{\op{n}_{i\sigma}}_0$. These equations reduce to Eq.~\ref{eq: Gutz-homo} in the homogeneous paramagnetic case, and to Eq.~\ref{eq: Gutz-y=1} in the square-lattice antiferromagnet. However, in a homogeneous partial ferromagnet, Eq.~\ref{eq: Gutz-Wang} is in disagreement with the results from Zhang \etal\cite{Zhang}, which are given by $g_t(\sigma)=(1-n_\up-n_\dn)/(1-n_\sigma)$ and $g_J = 1/(1-n_\up)(1-n_\dn)$. It should be noted that this $g_t$ obtained by Zhang \etal~is identical to that of Ogawa \etal,\cite{Ogawa} who did not derive $g_J$.

The brief survey above indicates that there has been confusion in what the appropriate Gutzwiller factors should be for inhomogeneous systems. The purpose of this paper is to clarify such confusion by \emph{deriving} the Gutzwiller factors by appropriately generalizing Ogawa's original approach,\cite{Ogawa} and to investigate the accuracy of the resulting Gutzwiller approximation in inhomogeneous systems by comparing with VMC results in two-dimensional square-lattice antiferromagnet (SLAF). 

Our derivation, which is presented in the appendix, is based on configuration counting on canonical (i.e., particle-number eigenstate) wavefunctions. This is a more natural choice for deriving the Gutzwiller factors, since the Gutzwiller approximation amounts to neglecting quantum correlations in configurations, \emph{but retaining combinatorial ones}. This combinatorial dependence is more clear in canonical wavefunctions as compared to grand-canonical ones (readers may want to compare our derivation with that of Ref.~\onlinecite{Wang}).

The key insight that resolves the confusion is that the single-particle density will in general be modified by the Gutzwiller projection, and that the relation between the projected density $\avg{\op{n}_{i\sigma}}$ and the pre-projected density $\avg{\op{n}_{i\sigma}}_0$ can be adjusted by introducing local fugacity factors $y_{i\sigma}$ that in general depend on both site and spin. This spin-and-site dependent fugacity factor was first introduced by Gebhard\cite{Gebhard} when deriving the Gutzwiller factors using a diagrammatic approach. Similar factors also appear in works related to Gutzwiller projected superconducting states,\cite{Edegger, Anderson, Laughlin} where the fugacity factor is introduced to keep the mean number of particles unchanged. It should be remarked that the spin dependence of $y_{i\sigma}$ is essential, since otherwise one can merely maintain $\avg{\op{n}_i} = \avg{\op{n}_i}_0$,  as in Ref.~\onlinecite{Wang}, but not $\avg{\op{n}_{i\sigma}} = \avg{\op{n}_{i\sigma}}_0$, and hence the local magnetization $\avg{\op{m}_i} = \avg{\op{n}_{i\up}}-\avg{\op{n}_{i\dn}}$ may be modified by the projection. When making the physical argument that the Gutzwiller factor is given by the probability for the process to occur in the projected state divided by the corresponding probability in the pre-projected state,\cite{Zhang} the renormalization of single-particle density must be taken into account. 

In the following sections, we shall find out that \emph{both} Eq.~\ref{eq: Gutz-adhoc} and Eq.~\ref{eq: Gutz-y=1} follow from the derivation, but correspond to two \emph{different implicit choices} of local fugacity factors. Specifically, the former results from demanding that $\avg{\op{n}_{j\sigma}}_0 = \avg{\op{n}_{j\sigma}}$ for all sites $j$ and for all spin $\sigma$, while the latter results from setting $y_{j,\sigma} = 1$ for all $j$ and $\sigma$. In the particular case of homogeneous paramagnet, our results reduce to Eq.~\ref{eq: Gutz-homo} \emph{regardless of the value of $y$}, which by symmetry is spin and site independent. This is expected for a canonical wavefunction and is in contrast with the derivation by Wang \etal,\cite{Wang} which demands a particular value of $y$ for the equations to work out. An understanding of this implicit choice of fugacities is particularly important if we want to compare results from Gutzwiller approximation to that from VMC calculations, since we need to make sure that we are comparing expectations with respect to the same wavefunction. 

We shall also discover subtleties associated with the definition of $g_J$ in section \ref{sect: intuitive}. In particular, the Gutzwiller factor $g_{Jz}$ for the $z$-component of the spin-spin interaction is in general different from the corresponding Gutzwiller factor $g_{Jxy}$ of the $x$- and $y$-component. The physical origin of this difference is that in a fixed spin basis, the Pauli exclusion principle posts a restriction on the legitimate configurations in the pre-projected wavefunction for the exchange of opposite spins, while such restriction is absent for exchange of same type of spins. Consequently, the Gutzwiller approximation \emph{breaks rotational symmetry} even in the homogeneous case, and Eq.~\ref{eq: Gutz-homo} can only be reached by enforcing rotational symmetry beyond the probability ratio argument.

It should be noted that the configuration counting approach we used here is in general different from the $1/d$-expansion developed by Metzner, Vollhardt and Gebhard.\cite{Gebhard, Metzner} It is however the configuration counting approach that corresponds to the physical intuition that the Gutzwiller factors are obtained by dividing the probability that a process would occur in the projected wavefunction by the corresponding probability in the pre-projected wavefunction (c.f. Sect.~\ref{sect: intuitive} and Appendix). In the case of SLAF, the formula for the various $g_J$ given by the two formulations are different, even when the $d\rightarrow\infty$ limit is taken in the $1/d$-expansion. However, the numerics agree qualitatively, and the numerical differences between the two approaches is of the same order of magnitude as the error between each approximation and the VMC result (c.f. Fig.~\ref{fig: Eplots2} in Sect.~\ref{sect: SLAF}).

\section{Intuitive Arguments for Gutzwiller Factors} \label{sect: intuitive}

\begin{table*}
\caption{Probability for various physical processes to occur in hopping}
{\small
\begin{ruledtabular}
\begin{tabular}{lllll}
Physical process &
\includegraphics[scale=0.4]{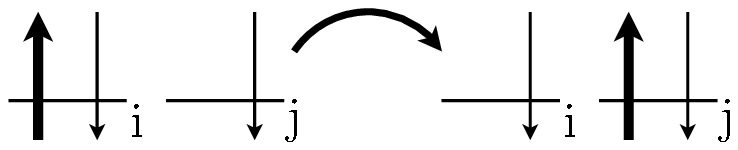} &
\includegraphics[scale=0.4]{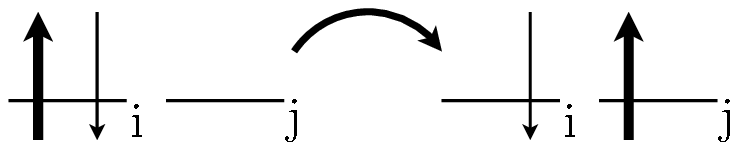} &
\includegraphics[scale=0.4]{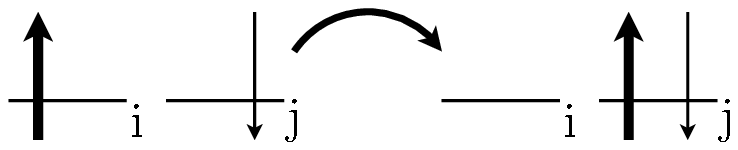} &
\includegraphics[scale=0.4]{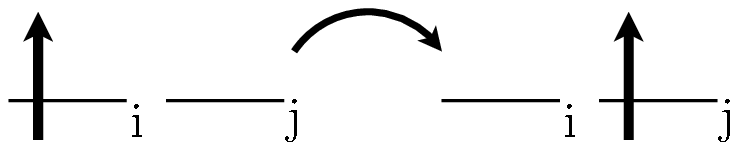} \\
\hline
Probability in $\psi_0$&
$\avg{\op{n}_{i\up}}_0 \avg{\op{n}_{i\dn}}_0$  &
$\avg{\op{n}_{i\up}}_0 \avg{\op{n}_{i\dn}}_0$  &
$\avg{\op{n}_{i\up}}_0 \avg{1-\op{n}_{i\dn}}_0$  &
$\avg{\op{n}_{i\up}}_0 \avg{1-\op{n}_{i\dn}}_0$  \\
(before process) & 
$\times \avg{1-\op{n}_{j\up}}_0 \avg{\op{n}_{j\dn}}_0$ & 
$\times \avg{1-\op{n}_{j\up}}_0 \avg{1-\op{n}_{j\dn}}_0$ & 
$\times \avg{1-\op{n}_{j\up}}_0 \avg{\op{n}_{j\dn}}_0$ & 
$\times \avg{1-\op{n}_{j\up}}_0 \avg{1-\op{n}_{j\dn}}_0$ \\
Probability in $\psi_0$&
$\avg{1-\op{n}_{i\up}}_0 \avg{\op{n}_{i\dn}}_0$  &
$\avg{1-\op{n}_{i\up}}_0 \avg{\op{n}_{i\dn}}_0$  &
$\avg{1-\op{n}_{i\up}}_0 \avg{1-\op{n}_{i\dn}}_0$  &
$\avg{1-\op{n}_{i\up}}_0 \avg{1-\op{n}_{i\dn}}_0$  \\
(after process) & 
$\times \avg{\op{n}_{j\up}}_0 \avg{\op{n}_{j\dn}}_0$ & 
$\times \avg{\op{n}_{j\up}}_0 \avg{1-\op{n}_{j\dn}}_0$ & 
$\times \avg{\op{n}_{j\up}}_0 \avg{\op{n}_{j\dn}}_0$ & 
$\times \avg{\op{n}_{j\up}}_0 \avg{1-\op{n}_{j\dn}}_0$ \\
Probability in $\psi$ & & & & \\
(before process) & 
\raisebox{1.5ex}[0cm][0cm]{0} &
\raisebox{1.5ex}[0cm][0cm]{0} & 
\raisebox{1.5ex}[0cm][0cm]{0} & 
\raisebox{1.5ex}[0cm][0cm]{$\avg{\op{n}_{i\up}} \avg{1-\op{n}_{j\up}-\op{n}_{j\dn}}$ } \\
Probability in $\psi$ & & & & \\
(after process) & 
\raisebox{1.5ex}[0cm][0cm]{0} &
\raisebox{1.5ex}[0cm][0cm]{0} & 
\raisebox{1.5ex}[0cm][0cm]{0} & 
\raisebox{1.5ex}[0cm][0cm]{$\avg{1-\op{n}_{i\up}-\op{n}_{i\dn}} \avg{\op{n}_{j\up}}$}  \\
\end{tabular}
\end{ruledtabular}
}
\label{fig: t_table}
\end{table*}

In the homogeneous paramagnetic case, an intuitive picture of the Gutzwiller approximation is provided by Zhang \etal,\cite{Zhang} whose central result is that the Gutzwiller factors are given by the probability for the physical process under consideration to occur in the projected wavefunction divided by the probability for such process to occur in the pre-projected wavefunction. A careful derivation of the Gutzwiller factors from the approach of Ogawa \etal,\cite{Ogawa} which we relegate to the appendix, confirms this intuitive picture even in the inhomogeneous case, with the exception that one can no longer assume the projected density $\avg{\op{n}_{i\sigma}}$ to equal to the pre-projected density $\avg{\op{n}_{i\sigma}}_0$. In this section, we shall review the argument by Zhang \etal, from which we shall see that $g_{Jz} \neq g_{Jxy}$. We shall also supplement the result of Zhang \etal~by providing an intuitive picture of how $\avg{\op{n}_{i\sigma}}$ is computed in the inhomogeneous case. For notational simplicity, henceforth we denote $\avg{\op{n}_{i\sigma}}$ as $\rho_{i\sigma}$ and $\avg{\op{n}_{i\sigma}}_0$ as $n_{i\sigma}$.

We shall work with the generalized Gutzwiller projector $\gutz' = \prod_j y_{j\up}^{\op{n}_{j\up}} y_{j\dn}^{\op{n}_{j\dn}} (1-\op{n}_{j\up} \op{n}_{j\dn})$, where $y_{j\sigma}$ are positive parameters that depend on both site and spin. To avoid the complications from particle number renormalization,\cite{Edegger, Anderson, Laughlin} throughout this paper $\ket{\psi_0}$ is assumed to be a spin-definite canonical wavefunction (i.e., fixed number of up-spin and down-spin particles).

First consider the hopping $\avg{\opdag{c}_{j\sigma} \op{c}_{i\sigma}}$. For concreteness take $\sigma = \up$. In the pre-projected wavefunction, before hopping an up-spin must reside on site-i, and Pauli exclusion principle demands no up-spin on site-j. Since there is no occupation constraints, one or zero down-spin can reside on each site. After hopping, the down-spins are unaffected while the up-spin moves to site-j, leaving site-i with no up-spins. There are thus four legitimate configurations in the pre-projected wavefunction. However, as a result of occupation constraint, only one of these four configurations is allowed in the projected wavefunction, namely the one which both sites contain no down-spins (see Table.~\ref{fig: t_table} for illustration). Summing over the probabilities of legitimate configurations, taking the ratio, and include an overall square root, we have:
\begin{equation} \label{eq: hop}
g_t(\sigma, i,j) = \sqrt{\frac{\rho_{i\sigma} (1-\rho_{j\up}-\rho_{j\dn})}
	{n_{i\sigma} (1-n_{j\sigma})} 
	\frac{\rho_{j\sigma} (1-\rho_{i\up}-\rho_{i\dn})}
	{n_{j\sigma} (1-n_{i\sigma}) }}\\
\end{equation}
This result is also produced and discussed in Ref.~\onlinecite{EdeggerReview}.

Next consider the spin-spin interaction. In a fixed spin basis, the spin-spin interaction consists of four distinct types of physical processes, corresponding to the different ways of expanding $\op{\vv{S}}_i \cdot \op{\vv{S}}_j$ in terms of $\op{c}$ and $\opdag{c}$, and various ways of Wick-contracting the four-fermion expectations. These processes are:
\begin{enumerate}
\item Exchange of an up-spin with a down-spin 
	($\sim \avg{\opdag{c}_{j\up}\op{c}_{j\dn}\opdag{c}_{i\dn}\op{c}_{i\up}}$)
\item Counting of one up-spin and one down-spin 
	($\sim \avg{\opdag{c}_{j\dn}\op{c}_{j\dn}\opdag{c}_{i\up}\op{c}_{i\up}}$)
\item Exchange of two spins of the same species
	($\sim \avg{
		\contraction{}{\opdag{c}_{j\up}}{\op{c}_{j\up}\opdag{c}_{i\up}}{\op{c}_{i\up}}{}
		\contraction[2ex]{\opdag{c}_{j\up}}{\op{c}_{j\up}}{}{\opdag{c}_{i\up}}
		\opdag{c}_{j\up}\op{c}_{j\up}\opdag{c}_{i\up}\op{c}_{i\up}
	}$)
\item Counting of two spins of the same species
	($\sim \avg{
		\contraction{}{\opdag{c}_{j\up}}{}{\op{c}_{j\up}}
		\contraction{\opdag{c}_{j\up}\op{c}_{j\up}}{\opdag{c}_{i\up}}{}{\op{c}_{i\up}}
		\opdag{c}_{j\up}\op{c}_{j\up}\opdag{c}_{i\up}\op{c}_{i\up}
	}$)
\end{enumerate}
Note that the first type of processes contributes to the x- and y- component of the spin-spin interaction, while the remaining three contribute to the z- component.

\begin{figure}
\includegraphics[scale=0.6]{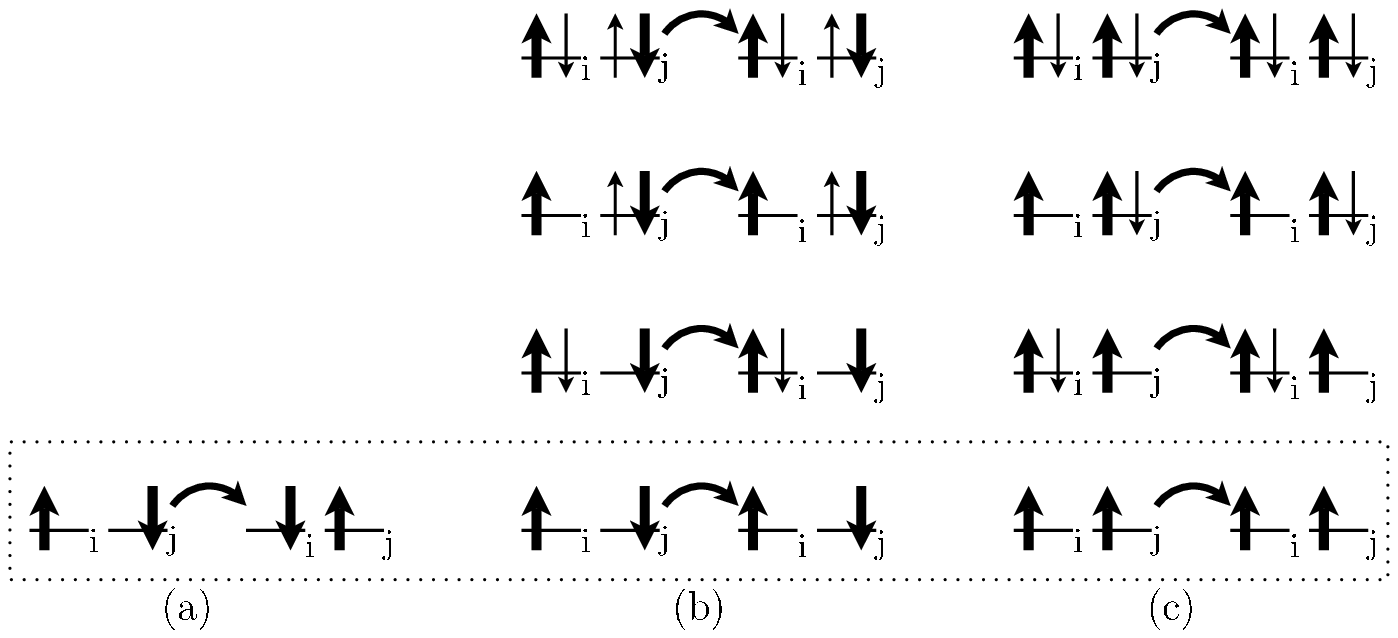}
\caption{Legitimate configurations in spin-spin interaction for the pre-projected wavefunction for various processes. Processes outside the rectangle in broken lines are disallowed in the projected wavefunction. (a) The process $\avg{\opdag{c}_{j\up}\op{c}_{j\dn}\opdag{c}_{i\dn}\op{c}_{i\up}}$ (type~1). (b) The process $\avg{\opdag{c}_{j\dn}\op{c}_{j\dn}\opdag{c}_{i\up}\op{c}_{i\up}}$ (type~2). (c) The processes in $\avg{\opdag{c}_{j\up}\op{c}_{j\up}\opdag{c}_{i\up}\op{c}_{i\up}}$ (types~3 and type~4).}
\label{fig: J_table}
\end{figure}

For processes of type~1, Pauli exclusion principle posts a strong restriction on the legitimate configurations in the \emph{pre-projected} wavefunction. In the specific example of $\avg{\opdag{c}_{i\up}\op{c}_{i\dn}\opdag{c}_{j\dn}\op{c}_{j\up}}$, it demands site-i to have no down-spin and site-j to have no up-spin. Consequently, the pre-projected wavefunction has only \emph{one} legitimate configuration for this process. In comparison, processes of types~2--4 all have four legitimate configurations in the pre-projected wavefunction. In other words, the occupation constraint is automatically demanded by processes of type~1, while has to be imposed additionally for process of types~2--4. See Fig.~\ref{fig: J_table} for illustration. 

Consequently, the probability for processes of type~1 to occur in the pre-projected wavefunction has an extra factor of $(1-n_{i\sigma})(1-n_{j\nn{\sigma}})(1-n_{i\nn{\sigma}})(1-n_{j\sigma})$ when compared with the other types of processes. Repeating the exercise of calculating the overall probability for a given type of process and group together various terms, the renormalization factors for the spin-spin interaction can be summarized as:
\begin{equation}\begin{split} \label{eq: Jfull}
\avg{\op{\vv{S}}_i \cdot \op{\vv{S}}_j} 
	= g_{Jxy} \avg{\op{S}_{i+} \op{S}_{j-} + \op{S}_{i-} \op{S}_{j+}}_0 
	+ g_{Jz} \avg{\op{S}_{iz}}_0 \avg{\op{S}_{jz}}_0 \\
	\qquad + \frac{g_{J\up}}{4} \avg{\opdag{c}_{j\up}\op{c}_{i\up}}_0 \avg{\op{c}_{j\up}\opdag{c}_{i\up}}_0 
	+ \frac{g_{J\dn}}{4} \avg{\opdag{c}_{j\dn}\op{c}_{i\dn}}_0 \avg{\op{c}_{j\dn}\opdag{c}_{i\dn}}_0
\end{split}\end{equation}
where,
\begin{align}
g_{Jxy} & = \sqrt{\frac{\rho_{i\up} \, \rho_{j\dn}}
	{n_{i\up} (1-n_{i\dn})	n_{j\dn} (1-n_{j\up})}
	\frac{\rho_{i\dn} \, \rho_{j\up}}
	{n_{i\dn} (1-n_{i\up}) n_{j\up} (1-n_{j\dn})}} \label{eq: Jxy}\\
g_{Jz} & = \frac{(\rho_{i\up}-\rho_{i\dn})(\rho_{j\up}-\rho_{j\dn})}
		{(n_{i\up}-n_{i\dn})(n_{j\up}-n_{j\dn})} \label{eq: Jz}\\
g_{J\up} & = \frac{\rho_{i\up} \rho_{j\up}}{n_{i\up} n_{j\up}} \label{eq: Jup}\\
g_{J\dn} & = \frac{\rho_{i\dn} \rho_{j\dn}}{n_{i\dn} n_{j\dn}} \label{eq: Jdn}
\end{align}
Note that the last three terms in Eq.~\ref{eq: Jfull} are all contributions from the z-component of the spin-spin interaction, while the first term is the contribution from the x- and y-component.

The four Gutzwiller factors above are in general unequal. In the specific case of a homogeneous paramagnet, $\rho_{i\sigma} = n_{i\sigma}$ and hence $g_{Jz} = g_{J\up} = g_{J\dn} = 1$ while $g_{Jxy} = 4/(1+\delta)^2$. However, since the Gutzwiller wavefunction for a homogeneous paramagnet is rotationally invariant, physical reasoning demands $\avg{\op{S}_{iz} \op{S}_{jz}} \approx g_{J\up} \avg{\opdag{c}_{j\up}\op{c}_{i\up}}_0 \avg{\op{c}_{j\up}\opdag{c}_{i\up}}_0/4 	+ g_{J\dn} \avg{\opdag{c}_{j\dn}\op{c}_{i\dn}}_0 \avg{\op{c}_{j\dn}\opdag{c}_{i\dn}}_0/4$ to equal to $\avg{\op{S}_{ix} \op{S}_{jx}} \approx g_{Jxy} \avg{\op{S}_{i+} \op{S}_{j-} + \op{S}_{i-} \op{S}_{j+}}_0 /2$. As the pre-projected wavefunction is also rotationally invariant, we must also have $\avg{\op{S}_{iz} \op{S}_{jz}}_0 = \avg{\op{S}_{ix} \op{S}_{jx}}_0$. Since $\avg{\op{S}_{iz} \op{S}_{jz}}_0$ is simply $\avg{\op{S}_{iz} \op{S}_{jz}}$ with all Gutzwiller factors set to 1, and similarly for $\avg{\op{S}_{ix} \op{S}_{jx}}_0$, one should expect $g_{J\up} = g_{J\dn} = g_{Jxy}$ in the homogeneous case. In other words, the Gutzwiller approximation derived from configuration counting breaks rotational symmetry. In retrospect this is not all surprising---the counting of configurations and the computation of probabilities are all done assuming a specific spin basis, and hence there is no {\itshape a priori\/} reason to expect rotational symmetry to be preserved. 

When applying the Gutzwiller approximation to near-homogeneous system, it may be desirable to set $g_{J\up} = g_{J\dn} = g_{Jxy}$ ``by hand.'' However, since in the homogeneous case $\avg{\op{S}_{iz}}_0 = 0$, we cannot similarly argue for setting $g_{Jz} = g_{Jxy}$. 

To compute the Gutzwiller factors, we still need to relate $\rho_{i\sigma}$ to $n_{i\sigma}$. Again we relegate the more mathematical derivation to the appendix and focus here on physical intuitions. Given any pre-projected spin-definite canonical wavefunction $\ket{\psi_0}$, we may expand $\ket{\psi_0}$ in the configuration basis. Those configurations with sites occupied by both up-spin and down-spin will be projected away by the Gutzwiller projector $\gutz'$. Since the Gutzwiller approximation conceptually amounts to neglecting correlation between different sites, each configuration that survives after projection should be assigned a classical weight $W$, which is a product of weighing factor $w(i)$ on each site i. The appropriate expression for $w(i)$ turns out to be:
\begin{equation}
w(i) = \left\{ \begin{array}{ll}
n_{i\sigma} (1-n_{i\nn{\sigma}}) y_{i\sigma}^2 & \textrm{for site occupied by spin $\sigma$}\\
(1-n_{i\up}) (1-n_{i\dn}) & \textrm{for empty site}\\
\end{array} \right.
\end{equation}
The above expression for $w(i)$ make conceptual sense: $n_{i\sigma} (1-n_{i\nn{\sigma}})$ and $(1-n_{i\up}) (1-n_{i\dn})$ are respectively the probabilities for finding a $\sigma$-spin and an empty site at site-i in the pre-projected wavefunction, and the factor of $y_{i\sigma}^2$ comes from the factor of $y_{i\sigma} ^{\op{n}_{i\sigma}}$ in both the bra and the ket of $\avg{\psi|\psi} = \bra{\psi_0} \gutz'^2 \ket{\psi_0}$.

After assigning each configuration with a classical weight, the projected density $\rho_{i\sigma}$ is simply given by the weighed average occupation of site-i by a $\sigma$-spin.

To derive an explicit set of equations that relate $\rho_{i\sigma}$ to $n_{i\sigma}$, assume that the system consists of $\kappa$ sublattices (labeled by $I$), such that on each sublattice $n_{i\sigma} = n_{j\sigma}$ and $y_{i\sigma} = y_{j\sigma}$. Let $N$ be the number of lattice site and $M$ = $N/\kappa$. Then, for a configuration with $a_{I\up}$ up-spins and $a_{I\dn}$ down-spins on sublattice $I$, the classical weight is:
\begin{equation} \begin{aligned} \label{eq: classical weight}
W(\{a_{I\sigma}\}) & = 
	\prod_I \big( (1-n_{I\dn}) (1-n_{I\up}) \big)^{M-a_{I\up}-a_{I\dn}} \\
	& \times \big( y_{I\up}^2 n_{I\up} (1-n_{I\dn}) \big)^{a_{I\up}}
	\big( y_{I\dn}^2 n_{I\dn} (1-n_{I\up}) \big)^{a_{I\dn}}
\end{aligned}\end{equation}

Moreover, simple combinatorics shows that the number of such configuration is given by:
\begin{equation} \label{eq: config count}
C(\{a_{I\sigma}\}) = \prod_I \frac{M!}{a_{I\up}! a_{I\dn}! (M-a_{I\up}-a_{I\dn})!}
\end{equation}

For a fixed site $i$ in sublattice $P$, a fraction of $a_{P\sigma}/M$ out of the $C(\{a_{I\sigma}\})$ configurations contain a $\sigma$-spin on site-i. Hence, the weighed average is given by:
\begin{equation} \label{eq: weighed avg}
\rho_{i \sigma} = \frac{
	\sum_{\{a_{I\sigma}\}} \! (a_{P\sigma}/M) 
	C(\{a_{I\sigma}\}) W(\{a_{I\sigma}\})}
	{\sum_{\{a_{I\sigma}\}} C(\{a_{I\sigma}\}) W(\{a_{I\sigma}\})}
\end{equation}

In the thermodynamic limit where $N \rightarrow \infty$, the function $F(\{a_{I\sigma}\}) = C(\{a_{I\sigma}\}) W(\{a_{I\sigma}\})$ is sharply peaked and hence the sum $\sum_{\{a_{I\sigma}\}}$ can be replaced by the single term in which $F(\{a_{I\sigma}\})$ attains maximum under the constraints $\sum_I a_{I\sigma} = \sum_j \bra{\psi_0} \op{n}_{j\sigma} \ket{\psi_0} = N_{\sigma}$. The maximizing $a_{I\sigma}$ can be found by standard Lagrange multiplier technique. Upon simplification, it is easy to check that $\rho_{i\sigma}$ can be solved through the following set of equations for a given choice of $y_{i\sigma}$:
\begin{equation} \label{eq: rho}
\left\{ \begin{aligned}
\frac{1-\rho_{I\up}-\rho_{I\dn}}{\rho_{I\sigma}}
	\frac{y_{I\sigma}^2 n_{I\sigma}}{1-n_{I\sigma}}
	= \lambda_\sigma & \qquad \forall I,\sigma\\
\sum_I \rho_{I\sigma} = \sum_I n_{I\sigma} & \qquad \forall \sigma &
\end{aligned} \right.
\end{equation}
here $\lambda_\sigma$ are Lagrange multiplier to be determined from the set of equations. Note that multiplying each $y_{I\up}$ by the same constant will not affect the equations relating $\rho_{I\sigma}$ to $n_{I\sigma}$, and similarly for multiplying the same constant to each $y_{I\dn}$. \footnote{The property that multiplying each $y_{I\up}$ (and similarly $y_{I\dn}$) by a constant does not affect the relationship between $\rho_{I\sigma}$ and $n_{I\sigma}$ holds strictly only for \emph{spin-definite canonical wavefunction} and is a consequence of its being an eigenstate of both $\op{N}_\up = \sum_i \op{n}_{i\up}$ and $\op{N}_\dn = \sum_i \op{n}_{i\dn}$. Consequently, Eq.~\ref{eq: rho} becomes approximate when the wavefunction is not spin definite or is grand-canonical, and can be inaccurate when, e.g., $\ket{\psi_0}$ is a highly smeared-out BCS state. c.f.  Ref.~\onlinecite{Edegger, Anderson, Laughlin}.}

With the choice $y_{I\sigma} = \sqrt{(1-n_{I\sigma})/(1-n_{I\up}-n_{I\dn})}$, we have $\avg{\op{n}_{i\sigma}} = \avg{\op{n}_{i\sigma}}_0$ \emph{within the Gutzwiller approximation}. With this implicit choice of $y_{I\sigma}$ then, we get:
\begin{align}
g_t(\sigma) & = \sqrt{\frac{(1-n_{i\up}-n_{i\dn})(1-n_{j\up}-n_{j\dn})}{(1-n_{i\sigma})(1-n_{j\sigma})}}\\
g_{Jxy} & = \frac{1}{\sqrt{(1-n_{i\up})(1-n_{i\dn})(1-n_{j\up})(1-n_{j\dn})}}\\
g_{Jz} & = g_{J\up} = g_{J\dn} = 1
\end{align}
If we assume $n_{i\up} = n_{i\dn} = \avg{\op{n}_i}_0/2$, we recover the $g_t$ and $g_{Jxy}$ quoted in Eq.~\ref{eq: Gutz-adhoc}. However, we now see that to be consistent with the approximation scheme, one needs to set $g_{Jz} = g_{J\up} = g_{J\dn} = 1$.

Note that the geometry and dimensionality of the underlying lattice have never entered the above discussion. This is a consequence of the approximations made in the configuration counting approach, where all quantum correlations except combinatorial ones are neglected. Combinatorial factors in general do not depend on the detailed geometry and dimensionality of the underlying lattice.

Since the results we obtained is independent of dimensionality, one may wonder if it is equivalent to the $d\rightarrow\infty$ limit of the $1/d$-expansion developed by Metzner, Vollhardt and Gebhard.\cite{Gebhard, Metzner} For the renormalization factor $g_t$, the approximations made by the two approaches are essentially the same, and the resulting factors are numerically equal.\cite{EdeggerReview} For the renormalization factor $g_J$, which involves a four-fermion expectation, the agreement is less spectacular. Although the derivation of Gebhard\cite{Gebhard} also indicates that $g_{Jz}$, $g_{Jxy}$, and $g_{J\sigma}$ should in general be different, in the case of SLAF the formula for $g_{J\sigma}$ from Gebhard's derivation disagrees with both $g_{J\sigma}$ and $g_{Jxy}$ we gave in Eq.~\ref{eq: Jxy}--\ref{eq: Jdn}. It should however be noted that, at least in the case of SLAF, the numerical energy estimates from the two formulations agree qualitatively, and the numerical differences between the two approaches is of the same order of magnitude as the error between each approximation and the VMC result (c.f. Fig.~\ref{fig: Eplots2} in Sect.~\ref{sect: SLAF}).

\section{The Case of Two-Dimensional SLAF} \label{sect: SLAF}

To determine the accuracy of the Gutzwiller approximation and various modification schemes of it that restore rotational invariance for the paramagnetic state, various expectation values in the two-dimensional SLAF are computed in these approximation schemes and are compared with the results obtained from VMC. In the following, we shall first treat $y_{i\sigma}$ as an additional variational parameter and consider the entire parameter space, from which we shall discover that there is a vast region in which the Gutzwiller approximation is erroneous. However, we shall also see that the expectation values depend strongly on physical density $\avg{\op{n}_{i\sigma}}$ and only weakly on how such density is obtained from the parameters. Consequently, there are paths within the full parameter space in which the full range of $\avg{\op{n}_{i\sigma}}$ can be explored and which the erroneous region can be avoided. We shall then focus on some of these paths and explicitly consider the different modification schemes.

For fixed hole density  $\delta = \sum_i (1-\op{n}_{i\dn} - \op{n}_{i\up})/N$, the pre-projected SLAF state is uniquely characterized by the antiferromagnetic order parameter $\Delta$.
Explicitly, the pre-projected wavefunction $\ket{\psi}$ is given by:
\begin{equation}
\ket{\psi} = \prod_{\epsilon_\vv{k} < \epsilon_F, \sigma} 
	(u_{\vv{k}} \opdag{c}_{\vv{k},\sigma} + 
	\func{sign}{\sigma} v_{\vv{k}} \opdag{c}_{\vv{k}+\vv{Q}, \sigma}) \ket{\emptyset}
\end{equation}
where $\vv{Q} = (\pi, \pi)$, $\epsilon_{\vv{k}} = - 2 (\cos k_x + \cos k_y)$, 
\begin{equation}\begin{aligned}
u_{\vv{k}}^2 & = \frac{1}{2}
	\left( 1 - \frac{\epsilon_{\vv{k}}}{\sqrt{\epsilon_{\vv{k}}^2 + \Delta^2}} \right) \\
v_{\vv{k}}^2 & = \frac{1}{2}
	\left( 1 + \frac{\epsilon_{\vv{k}}}{\sqrt{\epsilon_{\vv{k}}^2 + \Delta^2}} \right)
\end{aligned}\end{equation}
and which the Fermi energy $\epsilon_F$ is determined by the dopping $\delta$ via $\sum_{\sigma} \avg{\op{n}_{i\sigma}}_0 = 1-\delta$.

This pre-projected wavefunction is invariant under the simultaneous exchange of sublattice and spin indices. To preserve this symmetry after projection, we demand $y_{A\sigma} = y_{B\nn{\sigma}}$, where $A$, $B$ label sublattices. Since multiplying all fugacity factors by a constant amounts only to an overall normalization of the resulting projected wavefunction, the projection $\gutz' = \prod_j y_{j\up}^{\op{n}_{j\up}} y_{j\dn}^{\op{n}_{j\dn}} (1-\op{n}_{j\up} \op{n}_{j\dn})$ in this case is uniquely characterized by $y_r = y_{B\up}/y_{A\up}$. Hence, the wavefunction $\ket{\psi}$ to be considered in this section is uniquely characterized by the physically adjustable $\delta$ and the two trial-wavefunction parameters $(y_r, \Delta)$. Furthermore, since the mapping $(y_r, \Delta) \mapsto (y_r^{-1}, -\Delta)$ amounts to inverting the roles of sublattices $A$ and $B$, we need to consider only the parameter space in which $\Delta \geq 0$. 

Unless otherwise stated, the explicit data we present are at dopping $\delta = 0.025$, and the VMC calculations are performed in a lattice consisting of $8 \times 10$ sites, with periodic boundary conditions. We shall first make the comparisons over the whole parameter space $(y_r, \Delta)$ and then focus on specific paths within the parameter space.

For two-dimensional SLAF, Eq.~\ref{eq: rho} gives:
\begin{equation}\begin{aligned} \label{eq: rho-SLAF}
\rho_{A\up} = \rho_{B\dn} & = \frac{ n_{+} (n_{+}+n_{-})(1-n_{-}) }
	{ y_r^2 n_{-} (1-n_{+}) + n_{+}(1-n_{-}) }\\
\rho_{B\up} = \rho_{A\dn} & = \frac{ y_r^2 n_{-} (n_{+}+n_{-})(1-n_{+}) }
	{ y_r^2 n_{-} (1-n_{+}) + n_{+}(1-n_{-}) }
\end{aligned}\end{equation}
where $n_+ = n_{A\up} = n_{B\dn}$ and $n_- = n_{A\dn} = n_{B\up}$ as in Eq.~\ref{eq: Gutz-y=1}. It should be remarked that by setting $y_r = 1$ and plugging into Eq.~\ref{eq: Jxy}--\ref{eq: Jdn}, we find $g_{Jz} = g_{Jxy}$ and recover the results in Eq.~\ref{eq: Gutz-y=1} for $g_t$ and $g_J$ ($= g_{Jz} = g_{Jxy}$), while $g_{J\up} = g_{J\dn} = (1-n_+)(1-n_-) g_J$.

\begin{figure}
\includegraphics[scale=0.4]{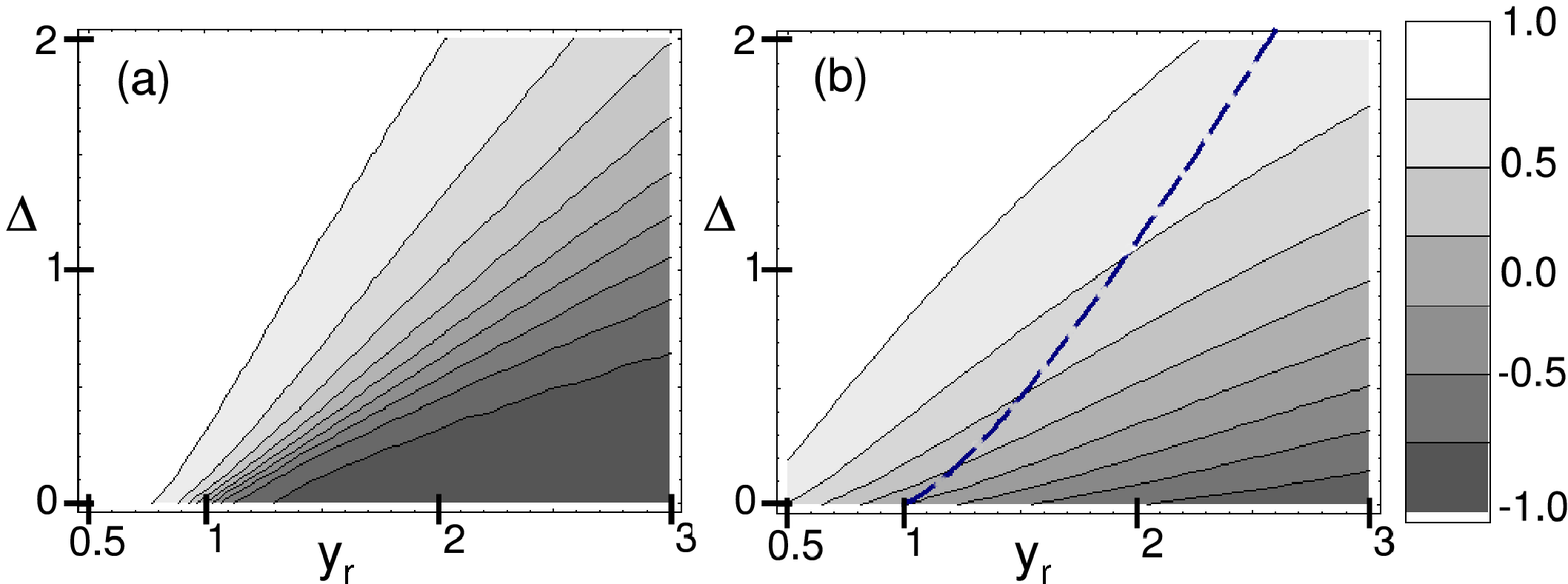} 
\caption{(Color online) Staggered magnetization $\avg{\op{m}}$ in (a) the VMC calculation and (b) the Gutzwiller approximation. The broken curve in (b) indicates the parameter subspace in which $\rho_{i\sigma} = n_{i\sigma}$ within the approximation. The VMC data is interpolated from a grid of interval $0.1$ in $y_r$ and $0.05$ in $\Delta$. The typical error in $\avg{\op{m}}$ in the VMC calculation is about $0.01$.}
\label{fig: mplots}
\end{figure}

From Eq.~\ref{eq: rho-SLAF} the staggered magnetization $\avg{\op{m}} = \rho_{A\up}-\rho_{A\dn}$ can immediately be evaluated. The results and the comparison with VMC results are shown in Fig.~\ref{fig: mplots}. Observe that constant magnetization contours from the Gutzwiller approximation are generally flatter than that from the VMC calculation.

\begin{figure}
	\subfigure[]{\includegraphics[scale=0.35]{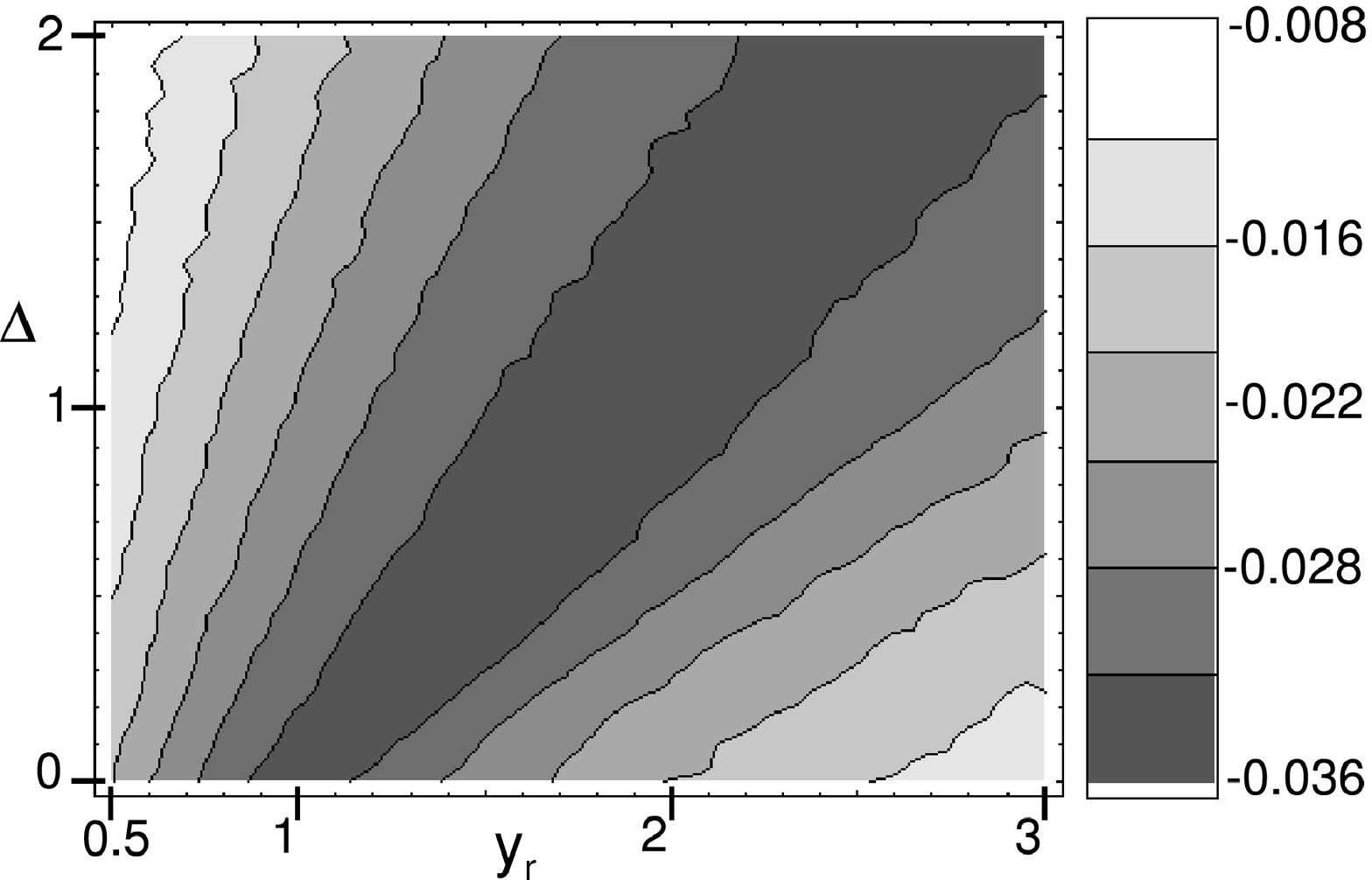}}
	\subfigure[]{\includegraphics[scale=0.35]{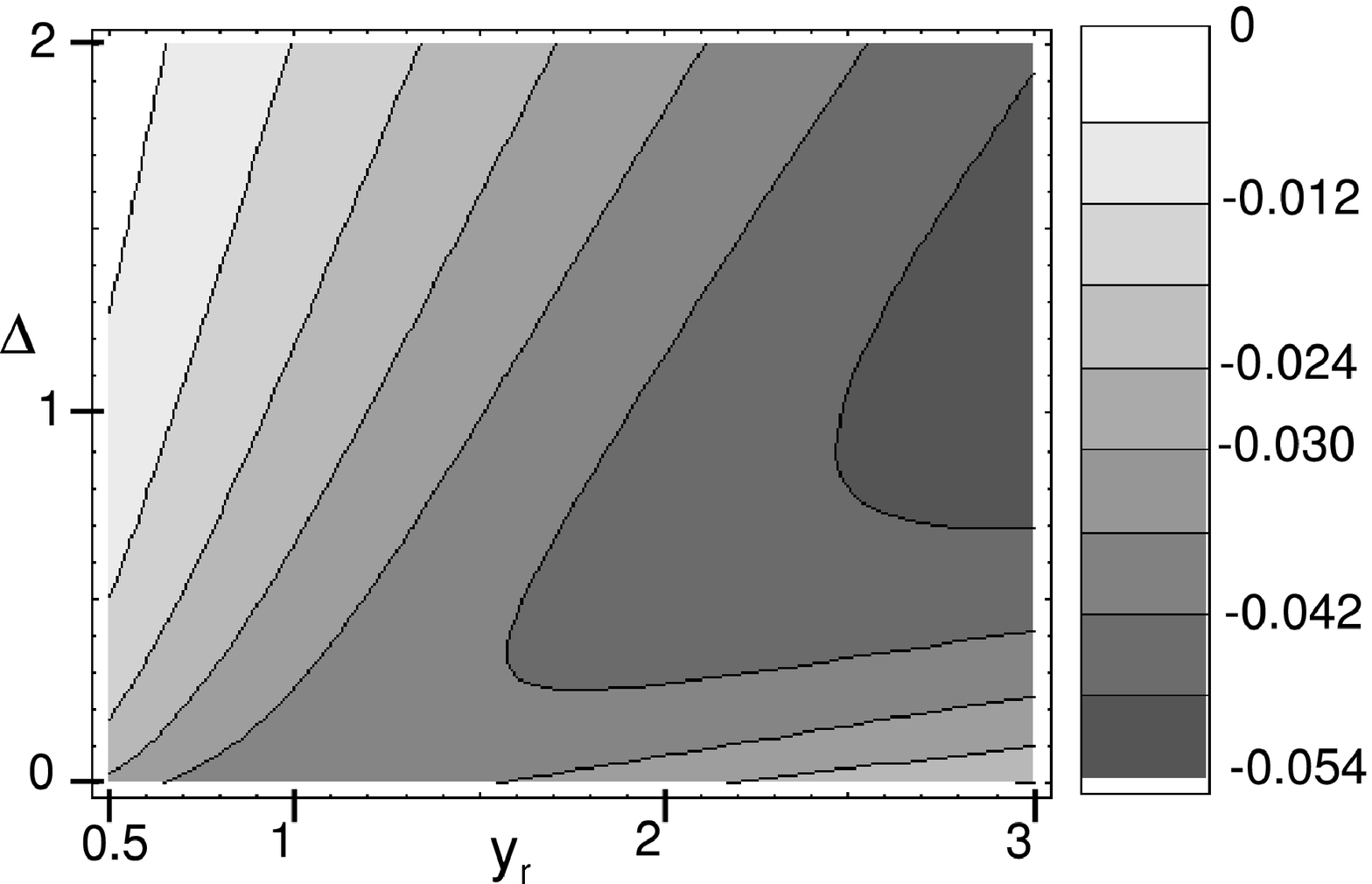}}
	\subfigure[]{\includegraphics[scale=0.35]{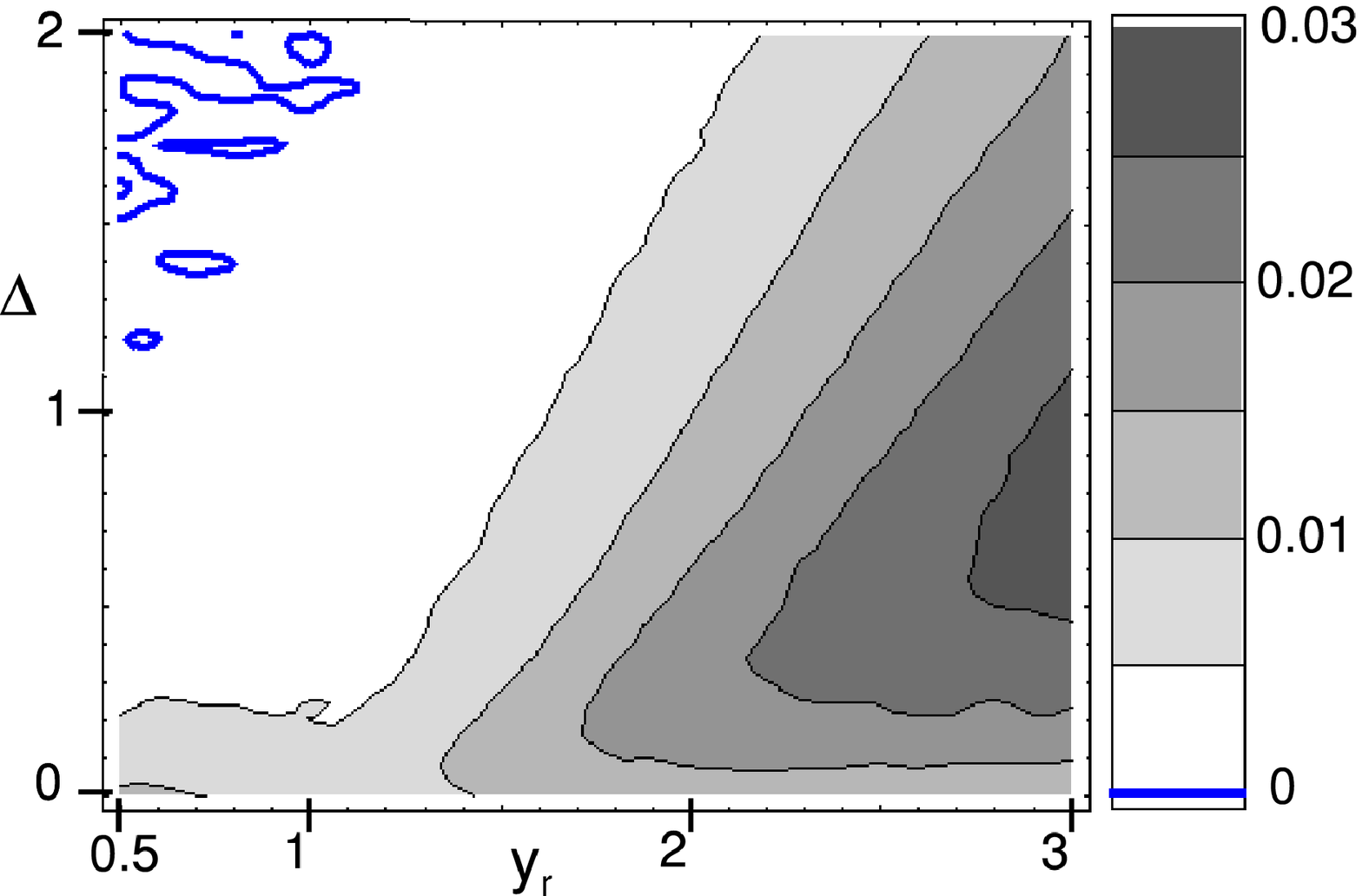}}
\caption{(Color online) Nearest neighbor hopping expectation $\avg{\opdag{c}_{i\sigma} \op{c}_{j\sigma}}$ in (a) the VMC calculation and (b) the Gutzwiller approximation, together with (c) the magnitude of difference between the two. The VMC data is interpolated from a grid of interval $0.1$ in $y_r$ and $0.05$ in $\Delta$. The typical error in $\avg{\opdag{c}_{i\sigma} \op{c}_{j\sigma}}$ in the VMC calculation is about $0.002$.} 
\label{fig: tplots}
\end{figure}

Next we consider the hopping expectation $\avg{\opdag{c}_{i\sigma} \op{c}_{j\sigma}}$ for nearest-neighbor sites $i,j$, which also gives $\avg{\Hn}$ in the $t$-model where $J=0$. From the properties of the wavefunction, this expectation is independent of $\sigma$ and the choice of particular sites. The results are shown in Fig.~\ref{fig: tplots}. It is important to note that while the hopping expectation calculated from VMC shows a shallow global minimum near $(y_r, \Delta) =(1,0)$, the hopping expectation calculated from the Gutzwiller approximation keeps decreasing along the direction in which both $y_r$ and $\Delta$ increase to maintain an approximately unchanged small staggered magnetization. \emph{This decrease is unphysical and indicates a systematic error in the Gutzwiller approximation} (as can be seen in Fig~\ref{fig: tplots}(c)). The origin of this high-error region in parameter space can be understood as follows: since the Gutzwiller approximation is based on neglecting non-combinatorial configuration dependences of expectation values, the effect of non-homogeneity caused by the fugacity factor $y_{i\sigma}$ (which is purely combinatorial) can be accurately accounted for in the approximation, while the effect of non-homogeneity caused by the parameters \emph{within} the pre-projected wavefunction $\ket{\psi_0}$ such as $\Delta$ cannot. The high-error region corresponds to trial wavefunctions in which \emph{each of} $y_{i\sigma}$ and $\Delta$ alone creates a large non-homogeneity, but which the two cancel out each other to produce an almost homogeneous state. In these states the effect of each of $y_{i\sigma}$ and $\Delta$ on $\avg{\opdag{c}_{i\sigma} \op{c}_{j\sigma}}$ is large, but the former is estimated accurately while the latter is estimated erroneously, thus producing a large overall error. Note also that the error in the Gutzwiller approximation becomes small when $\Delta \gg y_r$. This is not surprising as the wavefunctions become almost classical antiferromagnets in such limit.

\begin{figure}
	\subfigure[]{\includegraphics[scale=0.35]{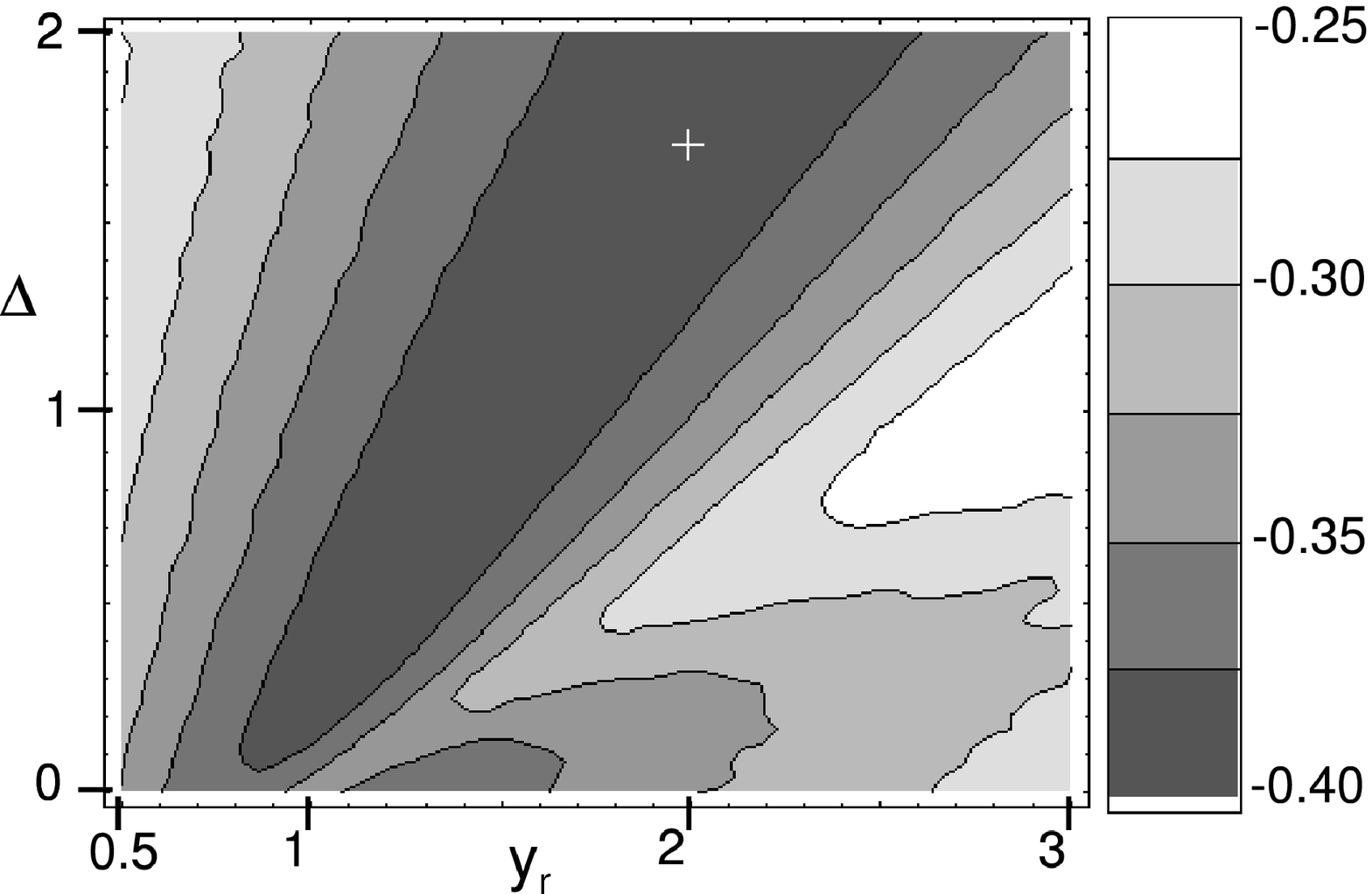}}
	\subfigure[]{\includegraphics[scale=0.35]{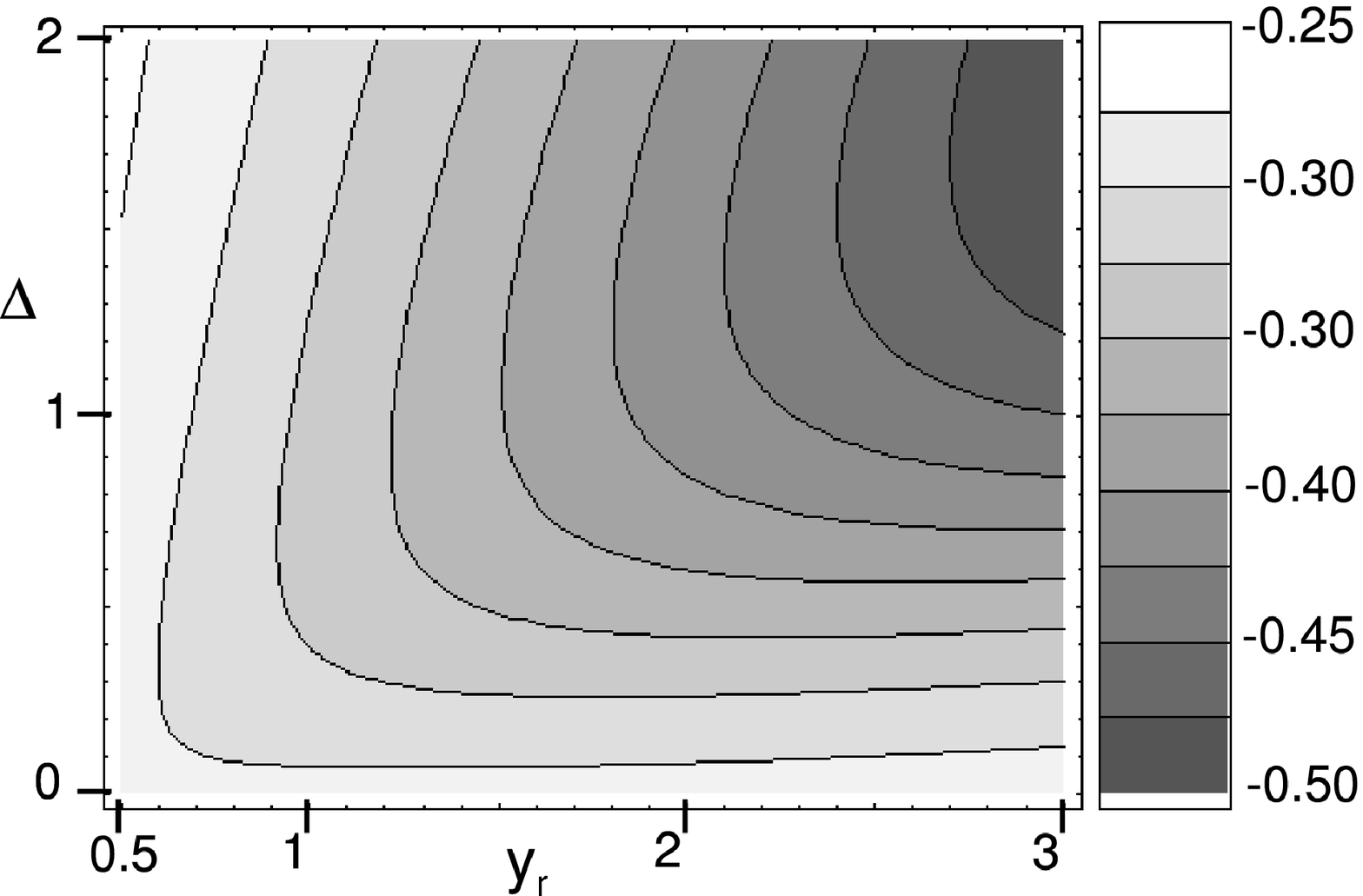}}
	\subfigure[]{\includegraphics[scale=0.35]{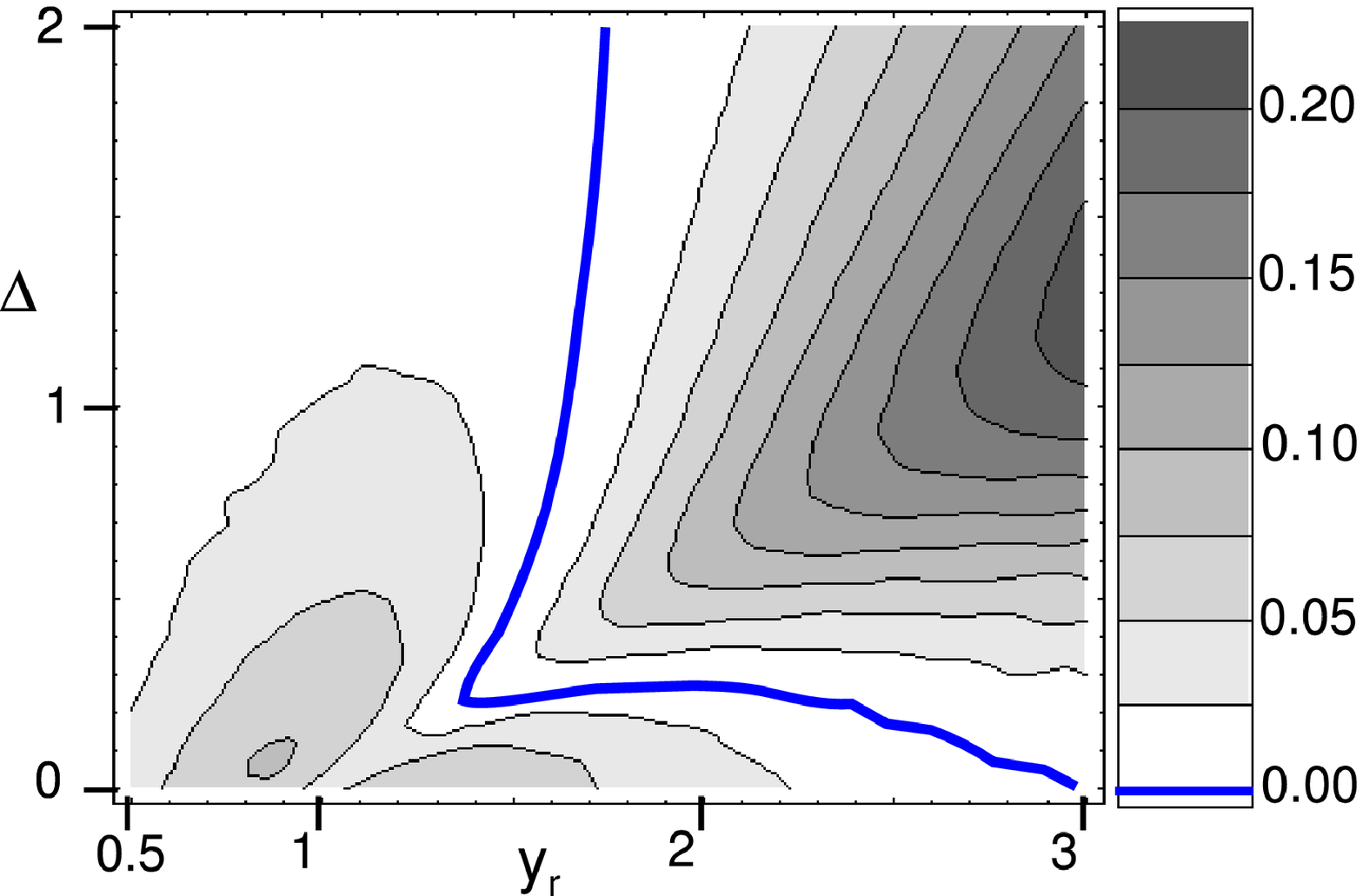}}
\caption{(Color online) Energy Expectation $\avg{\Hn}$, where $\Hn$ is defined as in Eq.~\ref{eq: Hamiltonian} with t=3 and J=1, in (a) the VMC calculation and (b) the Gutzwiller approximation, together with (c) the magnitude of difference between the two. The VMC data is interpolated from a grid of interval $0.1$ in $y_r$ and $0.05$ in $\Delta$. The typical error in $\avg{\Hn}$ in the VMC calculation is about $0.004$. The white cross in (a) indicates the location of global minimum in the VMC data set.}
\label{fig: Eplots}
\end{figure}

Next we consider the expectation of the Hamiltonian $\Hn$ as defined in Eq.~\ref{eq: Hamiltonian}, where we take the conventional ratio $t/J = 3$ and without the loss of generality set $J = 1$. The results are shown in Fig.~\ref{fig: Eplots}, where the Gutzwiller approximation of the $J$-term is implemented by Eq.~\ref{eq: Jxy}--\ref{eq: Jdn}. Again we see that whereas the expectation calculated from VMC shows a shallow global minimum, the expectation calculated from the Gutzwiller approximation keeps decreasing along a direction where both $y_r$ and $\Delta$ increase to maintain an approximately unchanged small staggered magnetization, which \emph{again indicates systemic error in the Gutzwiller approximation}. By splitting the contribution between the $t$-term and the $J$-term, it can be verified that this systemic error exists in the Gutzwiller approximation of both $\avg{\opdag{c}_{i\sigma} \op{c}_{j\sigma}}$ and  $\avg{\op{\vv{S}}_{i} \cdot \op{\vv{S}}_{j}}$. Moreover, whether we replace $g_{J\up}$ and $g_{J\dn}$ by $g_{Jxy}$ does not affect the existence of this systemic error.

Observe that in the VMC results, $\avg{\opdag{c}_{i\sigma} \op{c}_{j\sigma}}$ and $\avg{\Hn}$ along a constant staggered magnetization contour are approximately constant. It is thus sensible to speak of the various expectations as functions of $\avg{\op{n}_{i\sigma}}$, and to compute these quantities along a \emph{particular path} within the whole parameter space, as previous authors have implicitly done.\cite{Gan, Huang, Ziqiang, Poilblanc} \emph{Although the Gutzwiller approximation is inaccurate in certain regions of the whole parameter space, it may be acceptable along some specific paths.} In particular, the inaccurate region can be avoided by either paths with a fixed $y_r$ close to $1$ or by the path where $\rho_{i\sigma} = n_{i\sigma}$ within the Gutzwiller approximation. We shall now compare results from the Gutzwiller approximation and the VMC calculation along these paths, and explicitly consider various schemes that repair rotational symmetry. For paths of fixed $y_r$, we consider the following three schemes:
\begin{description}
\item[(z,z)] $g_{Jz}$, $g_{Jxy}$, $g_{J\up}$ and $g_{J\dn}$ as obtained from Eq.~\ref{eq: rho-SLAF} and Eq.~\ref{eq: Jxy}--\ref{eq: Jdn}.
\item[(z,xy)] $g_{Jz}$ and $g_{Jxy}$ as in scheme $(z,z)$, but with $g_{J\up} = g_{J\dn} = g_{Jxy}$.
\item[(xy,xy)] $g_{Jxy}$ as in scheme $(z,z)$, but with $g_{J\up} = g_{J\dn} = g_{Jz} = g_{Jxy}$.
\end{description}
For the path along which $\rho_{i\sigma} = n_{i\sigma}$ within the Gutzwiller approximation, we have one additional scheme:
\begin{description}
\item[diagrammatic] Gutzwiller factors obtained from diagrammatic approach,\cite{Gebhard} where $g_{Jz}$ and $g_{Jxy}$ agree with scheme (z,z), and which $g_{J\up}$ and $g_{J\dn}$ are given by:
\begin{equation} \label{eq: diagram}
g_{\up} = g_{\dn} = \frac{(1 - n_+ + n_-)(1 + n_+ - n_-)}{(1 - n_+)(1 - n_-)}
\end{equation}
\end{description}

Since we are interested in whether the Gutzwiller approximation can arrive at sensible \emph{physical} results, we shall compare between the Gutzwiller approximated expectations as functions of (projected) staggered magnetization \emph{within the approximation} and the VMC computed expectations as function of the VMC computed staggered magnetization.

\begin{figure}
	\subfigure[]{\includegraphics[scale=0.3, angle=0]{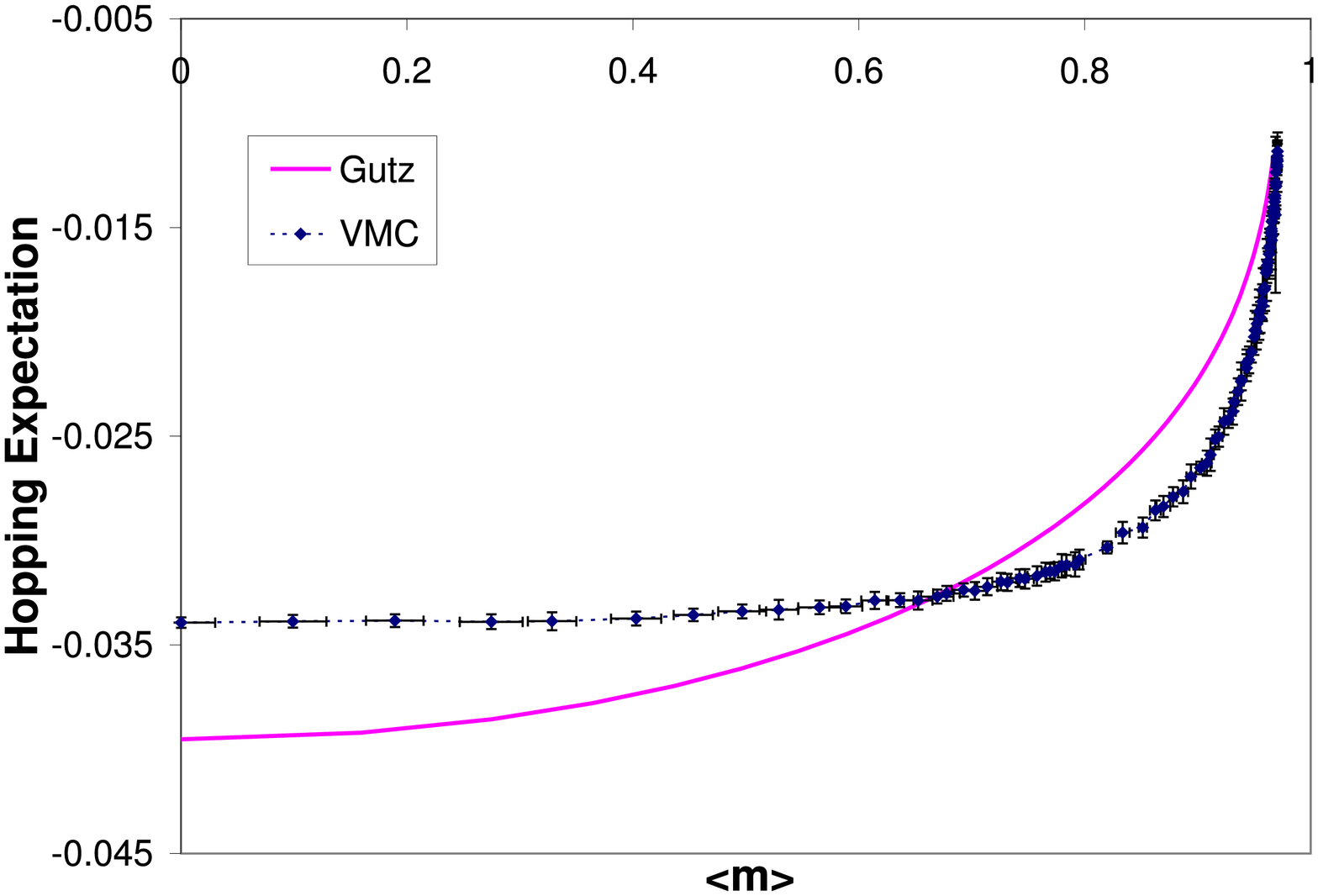}}
	\subfigure[]{\includegraphics[scale=0.3, angle=0]{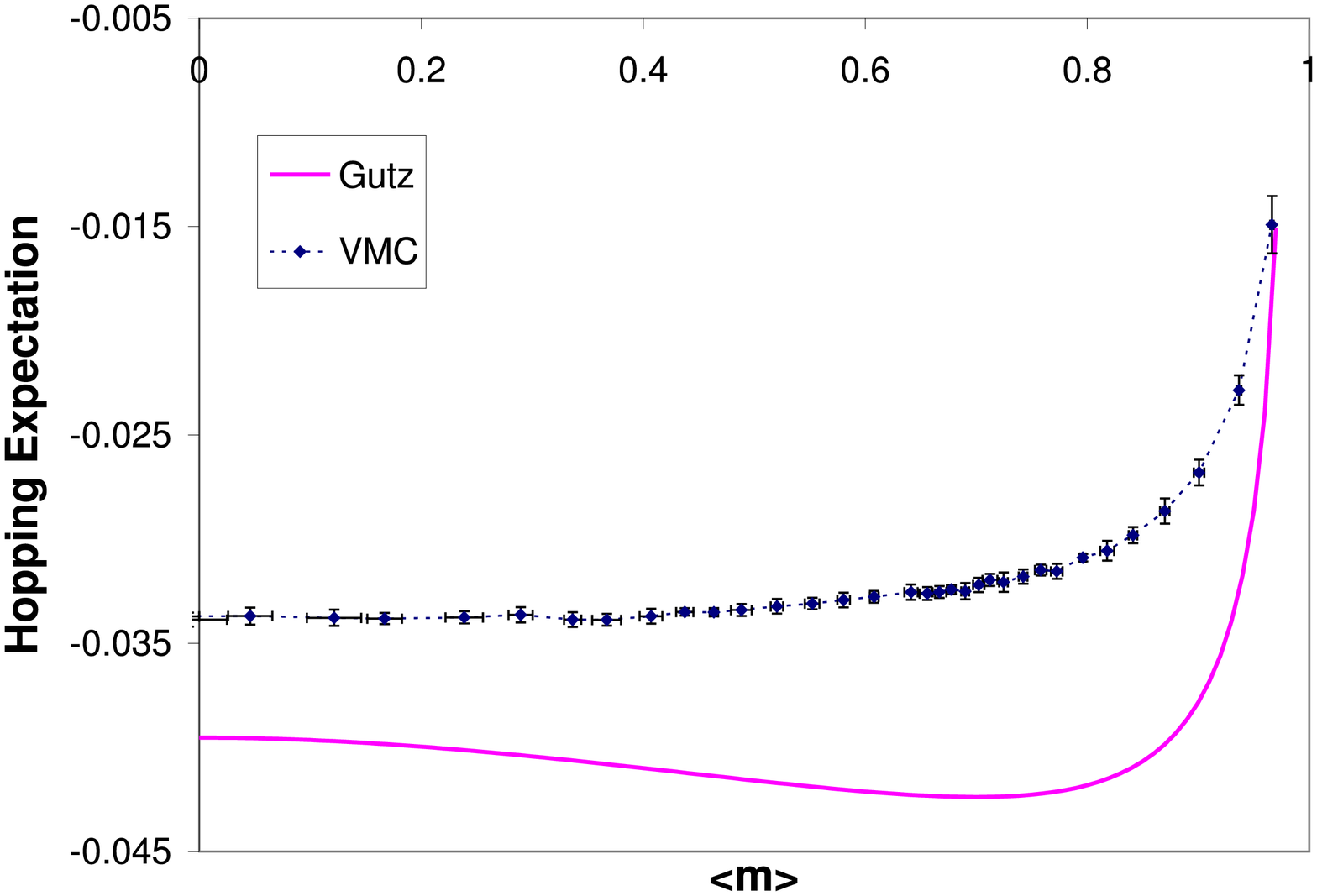}}
	\caption{(Color online) $\avg{\opdag{c}_{i\sigma} \op{c}_{j\sigma}}$ as function of $\avg{\op{m}}$ from the Gutzwiller approximation (unbroken purple line) and from the VMC calculation (blue filled dots), for (a) path along which $y_r = 1$ and (b) path along which $\rho_{i\sigma} = n_{i\sigma}$.}
	\label{fig: tplots2}
\end{figure}

First we consider the nearest-neighbor hopping expectation $\avg{\opdag{c}_{i\sigma} \op{c}_{j\sigma}}$, which relates trivially to the energy expectation in the t-model (i.e.\@ the t-J model with $J$ set to 0). The plots of $\avg{\opdag{c}_{i\sigma} \op{c}_{j\sigma}}$ as function of $\avg{\op{m}}$ along the path where $y_r = 1$ and $\rho_{i\sigma} = n_{i\sigma}$ are shown in Fig.~\ref{fig: tplots2}. The most striking observation from the plots comes from the Gutzwiller approximated result along $\rho_{i\sigma} = n_{i\sigma}$, which \emph{exhibits a minimum at $\avg{\op{m}} \approx 0.70$, contrary to the results indicated by the VMC calculation}. This false minimum is likely to be the remnant of the effect that produces the large error region, since the curve where $\rho_{i\sigma} = n_{i\sigma}$ is in close proximity with the large error region near $\avg{\op{m}} \approx 0.70$ (c.f.\@ fig.~\ref{fig: mplots} and \ref{fig: tplots}). Moreover, observe that the variation of $\avg{\opdag{c}_{i\sigma} \op{c}_{j\sigma}}$ as a function of $\avg{\op{m}}$ for small values of $\avg{\op{m}}$ in the VMC results is much smaller than the error between the VMC and the Gutzwiller results. Thus a small error in the Gutzwiller approximation for $\avg{\opdag{c}_{i\sigma} \op{c}_{j\sigma}}$ can lead to an incorrect prediction of the staggered magnetization at the t-model minimum-energy state. In other words, the systematic error in the Gutzwiller approximation may misled one into believing that an antiferromagnetic state is stabilized even in the t-model (see also the remark at the end of this section). In light of this, the result that spontaneous spin and charge ordering occurs in the triangular-lattice t-model\cite{triangular} should be viewed with some caution.

However, it should be noted that even without comparison with the VMC results, this inaccuracy could have at least been suspected, since the shape of $\avg{\opdag{c}_{i\sigma} \op{c}_{j\sigma}}$ as a function of $\avg{\op{m}}$ and the corresponding location of minima disagree between results from Gutzwiller approximation along the two different paths in the parameter space. In other words, \emph{checking the results for different paths within the parameter space may provide a consistency check within the Gutzwiller approximation}.

\begin{figure}
	\subfigure[]{\includegraphics[scale=0.32, angle=0]{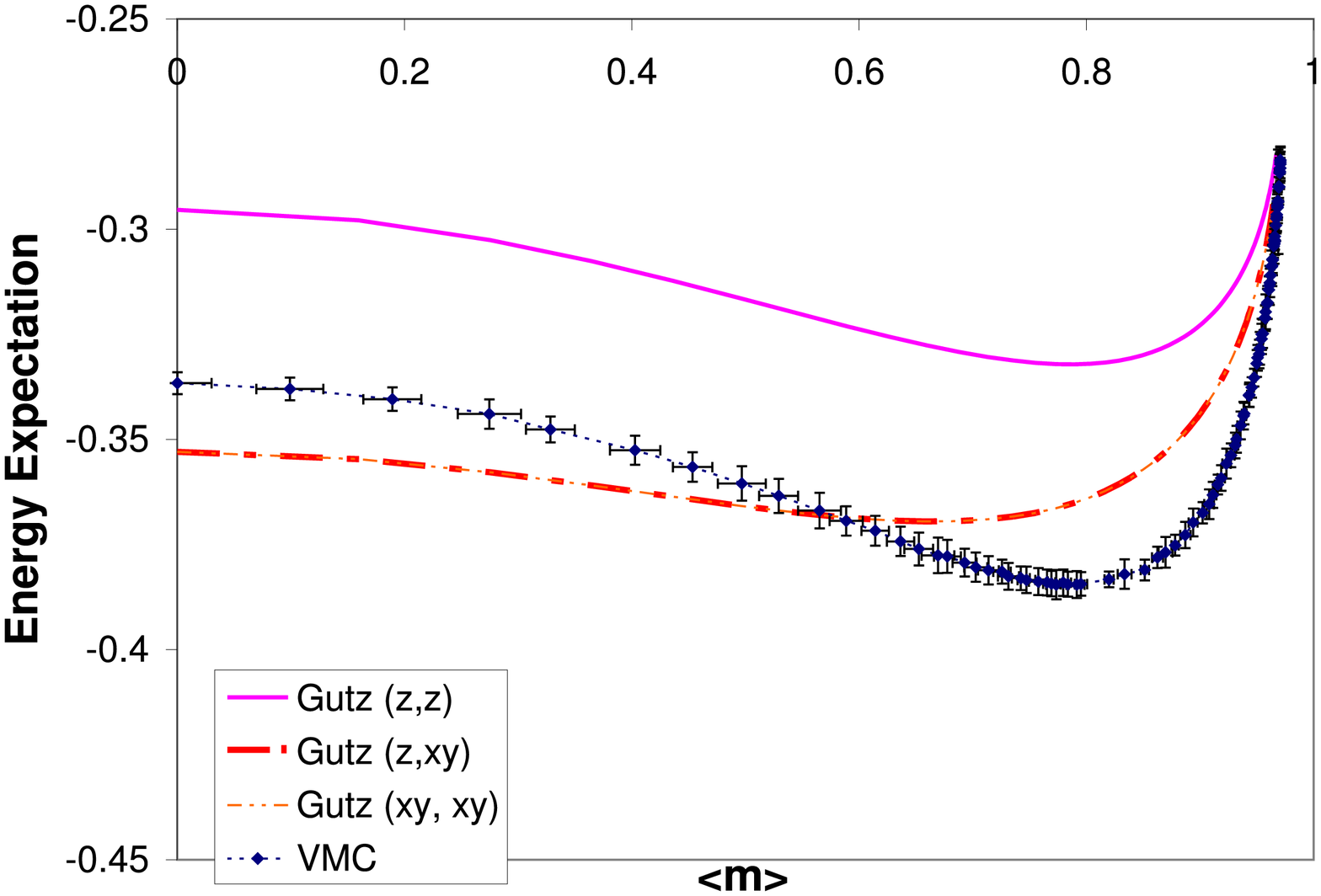}}
	\subfigure[]{\includegraphics[scale=0.32, angle=0]{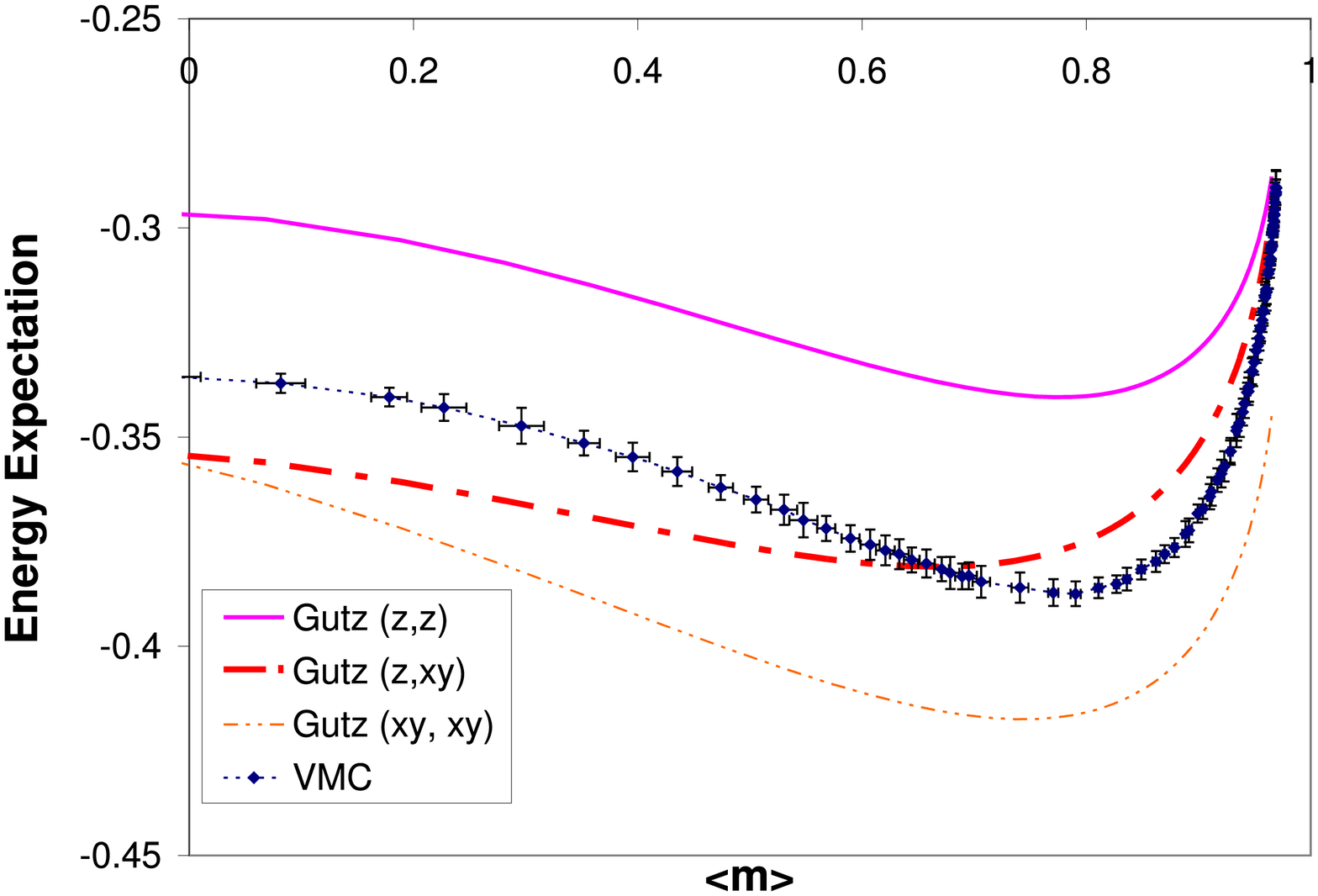}}
	\subfigure[]{\includegraphics[scale=0.32, angle=0]{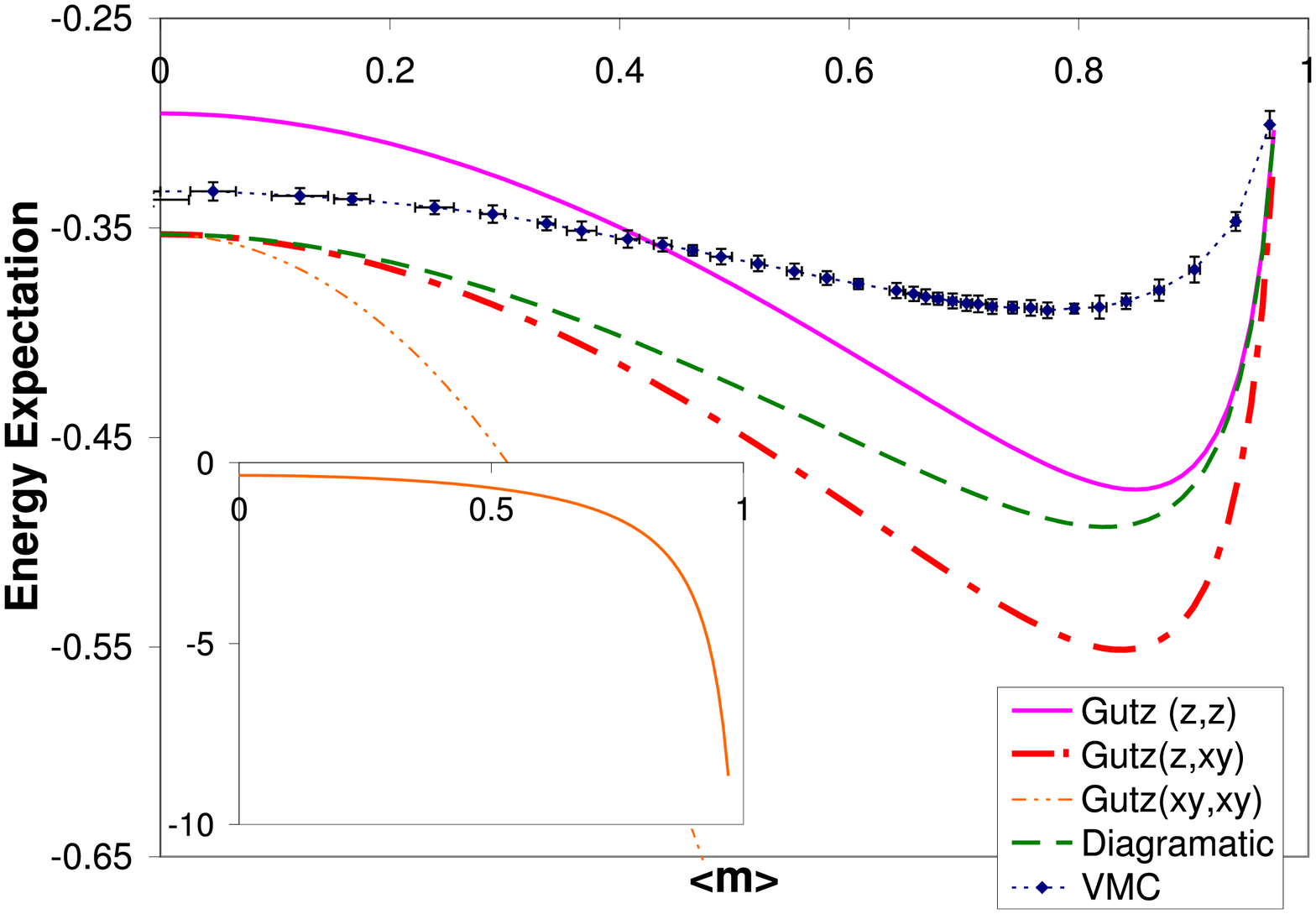}}
	\caption{(Color online) $\avg{\Hn}$ as function of $\avg{\op{m}}$ from the Gutzwiller approximation (with various schemes described in text) and from the VMC calculation, for (a) path along which $y_r = 1$, (b) path along which $y_r = 1.1$, and (c) path along which $\rho_{i\sigma} = n_{i\sigma}$. The inset of (c) plots the curve for scheme (xy,xy) in its full range.}
	\label{fig: Eplots2}
\end{figure}

Next consider the energy expectation $\avg{\Hn}$. The plots of $\avg{\Hn}$ as function of $\avg{\op{m}}$ along the paths where $y_r = 1$, $y_r = 1.1$ and $\rho_{i\sigma} = n_{i\sigma}$ are shown in Fig.~\ref{fig: Eplots2}. In this case, the VMC results consistently indicate a magnetized lowest-energy state around $\avg{\op{m}} \approx 0.78$. Moreover, the shapes of curves from the Gutzwiller approximation generally resemble the shapes of the VMC curves, with the notable exception of the (xy,xy) scheme along $\rho_{i\sigma} = n_{i\sigma}$. Recall that the rotational invariance for the paramagnetic state suggests merely that $g_{J\up} = g_{J\dn} = g_{Jxy}$, but bears no conclusion for $g_{Jz}$, since $\avg{\op{S}_{iz}}$ vanishes for such state. Combined with our numerical results here, \emph{it may be concluded that the (xy,xy) scheme should not be taken in a general Gutzwiller calculation.}\footnote{Even though results from scheme (z,xy) coincide with that from scheme (xy,xy) along the path $y_r = 1$, indicating that $g_{Jz} = g_{Jxy}$ along this path, the equality does not hold in general.} This conclusion is particularly important since the (xy,xy) scheme along $\rho_{i\sigma} = n_{i\sigma}$ is precisely the implicit scheme used by several authors.\cite{Huang, Ziqiang, Poilblanc} Among the remaining schemes, the (z,xy) and the diagrammatic schemes are probably preferred to the (z,z) scheme, since they provide better compromises between the accuracy in the small magnetization region and the large magnetization region.

We have also performed a similar comparison between the Gutzwiller and the VMC results at a higher dopping of $\delta = 0.125$, where the optimal magnetization for the t-J model decreases to approximately $0.40$. Most features of data discussed above in the $\delta = 0.025$ case continue to hold for the $\delta = 0.125$ case. The notable differences in the $\delta = 0.125$ case are: 1.\@ the erroneous minimum of $\avg{\opdag{c}_{i\sigma} \op{c}_{j\sigma}}$ as function of $\avg{\op{m}}$ in the Gutzwiller approximation disappears for the $\rho_{i\sigma} = n_{i\sigma}$ path; 2.\@ the VMC results show a slightly stronger dependence on the choice of particular paths in the parameter space; 3.\@ Results from the Gutzwiller approximation along the paths $y_r=1$ and $y_r = 1.1$ significantly underestimated the optimal magnetization for $\avg{\Hn}$ (even though the shape of the curves still resemble that of the VMC results). The second and third differences are possibly caused by the flatness of the minimum, which is an indication that this dopping is close to the critical value of antiferromagnet--paramagnet transition.

It should be remarked that the VMC calculation along the $y_r = 1$ line has been done previously by Yokoyama and Shiba\cite{Shiba} and our results are consistent with theirs. It should also be noted that we have restricted our consideration to the antiferromagnet states in order to investigate the accuracy of the Gutzwiller approximation and its various reparation schemes in a clear and simple inhomogeneous case. The antiferromagnetic state is somewhat artificial for the $t$-model at the dopping we considered, since the Nagaoka effect\cite{Nagaoka} suggests that the ground state should be ferromagnetic at low dopping, and indeed it can be checked that the ferromagnetic states have a lower expectation $\avg{\opdag{c}_{i\sigma} \op{c}_{j\sigma}}$ for both $\delta = 0.025$ and $\delta = 0.125$.\cite{Shiba}

\section{Summary}

In this paper we have considered generalizing the Gutzwiller approximation to the case of an inhomogeneous system and have found it useful to introduce extra spin-and-site-dependent fugacity factors. We derived the corresponding Gutzwiller factors from a configuration-counting approach in the appendix and explained its physical intuitions in the main text. The inclusion of fugacity factors reconcile the seemingly contradictory choices of Gutzwiller factors in the literature. Specifically, different Gutzwiller factors that appear in the literature correspond to different \emph{implicit} choices of fugacity factors. This fact is particularly important when comparing results from the Gutzwiller approximation to those from other approaches, such as variational Monte-Carlo. The derivation and discussion of the Gutzwiller factors also show that the Gutzwiller approximation generally breaks the rotational symmetry of the trial wavefunction. Specifically, different components of the spin-spin interaction are renormalized differently and  remain different even in the homogeneous case, contrary to the assertions from the bulk of the literature. We proposed several possible schemes to remedy this defect of the Gutzwiller approximation, some of which are implicitly applied by various authors in the literature. It should be noted that these remedies are, strictly speaking, outside the scope of the configuration-counting approach of the Gutzwiller approximation. To compare the accuracy of the Gutzwiller approximation from various choices of fugacity factors and various choices of reparation schemes, we perform calculations for the two-dimensional square-lattice antiferromagnet and compare with the results from variational Monte-Carlo. Stated in general terms, the ``lessons'' learnt from the comparison are:
\begin{enumerate}
\item There are regions in parameter space where the Gutzwiller approximation is erroneous. In particular, one should avoid the parameter space where the pre-projected wavefunction and the fugacity factors each alone produces large inhomogeneity, but which the two counteract each other strongly.
\item In general, one cannot assume $g_{Jxy}$ and $g_{Jz}$ (defined in Eq.~\ref{eq: Jfull}--\ref{eq: Jdn}) to be equal, which otherwise can lead to qualitatively and quantitatively erroneous results. It may be advisable, however, to take $g_{J\up} = g_{J\dn} = g_{Jxy}$ based on the consideration of rotational invariance.
\item  The energy expectation in both the $t$ model and $t$-$J$ model seems to depend strongly on physical parameters, namely the single-site spin-specific density $\avg{\op{n}_{i\sigma}}$, and weakly on whether such density is produced from the fugacity $y_{i\sigma}$ or from the pre-projected wavefunction $\ket{\psi_0}$. States that produce the same $\avg{\op{n}_{i\sigma}}$ with different $y_{i\sigma}$ thus provide a possible consistency check.
\end{enumerate}

In short, it is advisable to use the (z,xy) scheme defined above Eq.~\ref{eq: diagram} in calculations, and perform the calculations along at least two different paths in parameter space (conveniently the $y_r = 1$ and $\avg{\op{n}_{i\sigma}} = \avg{\op{n}_{i\sigma}}_0$ paths) for a consistency check.

\begin{acknowledgments}
We thank Ziqiang Wang and Sen Zhou for helpful discussions. We particularly thank Qiang-Hua Wang and Fu-Chun Zhang for helpful comments on our earlier manuscript. This work was supported by the Department of Energy under grant no.\@ DE-FG02-03ER46076.\\
\end{acknowledgments}

\appendix

\section{Exact Expressions for Expectations of Gutzwiller Projected Wavefunctions} 

We shall for simplicity assume the pre-projected wavefunction $\ket{\psi_0}$ to be normalized, canonical and spin-definite. Explicitly, let $\ket{\emptyset}$ denotes the vacuum state, and let $\op{\alpha}_{k\sigma}$ be linear combinations of $\op{c}_{j\sigma}$ such that $\bra{\emptyset} \op{\alpha}_k \opdag{\alpha}_{k'} \ket{\emptyset} = \delta_{k k'}$. Then, $\ket{\psi_0}$ is assumed to take the form:
\begin{equation} \label{eq: Slater}
\ket{\psi_0} = \prod_{k'} \opdag{\alpha}_{k'\up} \prod_k \opdag{\alpha}_{k\dn} \ket{\emptyset}, 
\end{equation}

We shall also assume that there is some superlattice $\{\vv{R}\}$ such that $\avg{\op{n}_{j,\sigma}}_0 = \avg{\op{n}_{j+\vv{R},\sigma}}_0$ and $y_{j,\sigma} = y_{j+\vv{R},\sigma}$.

Now, define $\op{G}_j$ on site-$j$ as:
\begin{equation} \label{eq: proj}
\begin{aligned} 
\op{G}_j & = y_{j\up}^{\op{n}_{j\up}} y_{j\dn}^{\op{n}_{j\dn}} (1-\op{n}_{j\up} \op{n}_{j\dn}) \\
& = (1-\op{n}_{j\up})(1-\op{n}_{j\dn}) + y_{j\up} \op{n}_{j\up}(1-\op{n}_{j\dn}) + y_{j\dn} \op{n}_{j\dn}(1-\op{n}_{j\up}) \\
& = \op{E}_{0j} + y_{j\up} \op{E}_{\up j} + y_{j\dn} \op{E}_{\dn j} 
\end{aligned}\end{equation}
where $\op{E}_{0j}$, $\op{E}_{\up j}$ and $\op{E}_{\dn j}$ are projection operators onto empty $j$-site, up-spin $j$-site, and down-spin $j$-site respectively. Note that these projection operators are orthogonal to each other (i.e., $\op{E}_{\alpha j} \op{E}_{\beta j} = \delta_{\alpha \beta} \op{E}_{\beta j}$), and that $\gutz' = \prod_j \op{G}_j$. 

It is then an easy exercise to show that $\op{G}_j \: \opdag{c}_{j\sigma} = y_{j\sigma} \opdag{c}_{j\sigma} 
	(\op{c}_{j\nn{\sigma}} \opdag{c}_{j\nn{\sigma}})$, and hence for distinct sites $i$,$j$,
\begin{equation}\begin{aligned}
\label{eq: GiiG}
	\op{G}_j \opdag{c}_{j\sigma} \op{c}_{j\sigma} \op{G}_j & = y_{j\sigma} \op{E}_{\sigma j} \\
	\op{G}_j \op{G}_i \opdag{c}_{j\sigma} \op{c}_{i\sigma} \op{G}_i \op{G}_j & =
	y_{j\sigma} y_{i\sigma} (\op{c}_{j\nn{\sigma}} \opdag{c}_{j\nn{\sigma}}) 
	(\opdag{c}_{j\sigma}\op{c}_{i\sigma}) (\op{c}_{i\nn{\sigma}} \opdag{c}_{i\nn{\sigma}})\\
	\op{G}_j \op{G}_i \opdag{c}_{j\sigma} \op{c}_{j\nn{\sigma}} \opdag{c}_{i\nn{\sigma}}
	\op{c}_{i\sigma} \op{G}_i \op{G}_j & = y_{i\up} y_{i\dn} y_{j\up} y_{j\dn}
	\opdag{c}_{j\sigma} \op{c}_{j\nn{\sigma}} \opdag{c}_{i\nn{\sigma}} \op{c}_{i\sigma}
\end{aligned}\end{equation}

From the orthogonality of the projection operators in Eq.~\ref{eq: proj} and the explicit form of wavefunction assumed in Eq.~\ref{eq: Slater}, we have:

\begin{widetext}
\begin{equation}\begin{aligned} \label{eq: norm_raw}
\avg{\psi|\psi} & = \bra{\psi_0} \prod_j \op{G}_j^2 \ket{\psi_0}
		= \sum_{ (\set{A},\set{B}) } \bra{\psi_0} 
		\prod_{j\in\set{A}} y_{j\up}^2 \op{E}_{\up j} 
		\prod_{j\in\set{B}} y_{j\dn}^2 \op{E}_{\dn j} 
		\nquad \prod_{j\notin \set{A} \cup \set{B}} \op{E}_{0 j} \ket{\psi_0}\\
	& = \sum_{ (\set{A},\set{B}) } \left( \prod_{j\in\set{A}} y_{j\up}^2 \right) 
		\left( \prod_{j\in\set{B}} y_{j\dn}^2 \right)
		\bra{\psi_0^\up} \prod_{j\in\set{A}} \op{n}_{j \up}
		\prod_{j\notin\set{A}} (1-\op{n}_{j \up}) \ket{\psi_0^\up}
		\bra{\psi_0^\dn} \prod_{j\in\set{B}} \op{n}_{j \dn}
		\prod_{j\notin\set{B}} (1-\op{n}_{j \dn}) \ket{\psi_0^\dn}
\end{aligned}\end{equation}
Here $\ket{\psi_0^\sigma} = \prod_k \opdag{\alpha}_{k\sigma} \ket{\emptyset}$, with $\opdag{\alpha}_{k\sigma}$ defined as in Eq.~\ref{eq: Slater}, and the sum is over all subsets $\set{A}, \set{B}$ of the set of lattice sites $\set{L}$ such that $|\set{A}| = \sum_j \bra{\psi_0} \op{n}_{j\up}\ket{\psi_0} = N_{\up}$, $|\set{B}| = \sum_j \bra{\psi_0} \op{n}_{j\dn}\ket{\psi_0} = N_{\dn}$, and $\set{A} \cap \set{B} = \emptyset$.

Taking also Eq.~\ref{eq: GiiG} into account, we have, analogously,
\begin{equation}\begin{aligned} \label{eq: density_raw}
\bra{\psi} & \opdag{c}_{i\up} \op{c}_{i\up} \ket{\psi} 
	= \sum_{\substack{(\set{A},\set{B}) \\ i\in \set{A}}} 
		\left( \prod_{j\in\set{A}} y_{j\up}^2 \right) 
		\left( \prod_{j\in\set{B}} y_{j\dn}^2 \right)
		\bra{\psi_0^\up} \prod_{j\in\set{A}} \op{n}_{j \up}
		\prod_{j\notin\set{A}} (1-\op{n}_{j \up}) \ket{\psi_0^\up}
		\bra{\psi_0^\dn} \prod_{j\in\set{B}} \op{n}_{j \dn}
		\prod_{j\notin\set{B}} (1-\op{n}_{j \dn}) \ket{\psi_0^\dn}
\end{aligned}\end{equation}
\begin{equation}\begin{aligned} \label{eq: hop_raw}
\bra{\psi} & \opdag{c}_{j\up} \op{c}_{i\up} \ket{\psi} 
	= \nquad \sum_{\substack{ (\set{A},\set{B}) \\ i\in \set{A}, j\notin \set{A}\cup\set{B}}} 
		\nquad y_{i\up} y_{j\up} \left( \prod_{\ell\in\set{A}\setminus\{i\}} y_{\ell\up}^2 \right)
		\left( \prod_{\ell\in\set{B}} y_{\ell\dn}^2 \right)
		\bra{\psi_0^\up} \opdag{c}_{j\up} \op{c}_{i\up}
		\nquad \prod_{\ell\in\set{A}\setminus\{i\}} \nquad \op{n}_{\ell\up}
		\!\! \prod_{\ell\notin\set{A}\cup\{j\}} \nquad (1-\op{n}_{\ell\up}) \ket{\psi_0^\up}
		\bra{\psi_0^\dn} \prod_{\ell\in\set{B}} \op{n}_{\ell\dn}
		\prod_{\ell\notin\set{B}} (1-\op{n}_{\ell \dn}) \ket{\psi_0^\dn}
\end{aligned}\end{equation}
\begin{equation}\begin{aligned} \label{eq: Jz1_raw}
\bra{\psi} & \opdag{c}_{i\up} \op{c}_{i\up} \opdag{c}_{j\up} \op{c}_{j\up} \ket{\psi} 
	= \sum_{\substack{(\set{A},\set{B}) \\ i,j\in \set{A}}}
		\left( \prod_{\ell\in\set{A}} y_{\ell\up}^2 \right)
		\left( \prod_{\ell\in\set{B}} y_{\ell\dn}^2 \right)
		\bra{\psi_0^\up} \prod_{\ell\in\set{A}} \op{n}_{\ell\up}
		\prod_{\ell\notin\set{A}} (1-\op{n}_{\ell\up}) \ket{\psi_0^\up}
		\bra{\psi_0^\dn} \prod_{\ell\in\set{B}} \op{n}_{\ell\dn}
		\prod_{\ell\notin\set{B}} (1-\op{n}_{\ell\dn}) \ket{\psi_0^\dn}
\end{aligned}\end{equation}
\begin{equation}\begin{aligned} \label{eq: Jz2_raw}
\bra{\psi} & \opdag{c}_{i\up} \op{c}_{i\up} \opdag{c}_{j\dn} \op{c}_{j\dn} \ket{\psi} 
	= \sum_{\substack{(\set{A},\set{B}) \\ i\in \set{A}, j\in \set{B}}}
		\left( \prod_{\ell\in\set{A}} y_{\ell\up}^2 \right)
		\left( \prod_{\ell\in\set{B}} y_{\ell\dn}^2 \right)
		\bra{\psi_0^\up} \prod_{\ell\in\set{A}} \op{n}_{\ell\up}
		\prod_{\ell\notin\set{A}} (1-\op{n}_{\ell\up}) \ket{\psi_0^\up}
		\bra{\psi_0^\dn} \prod_{\ell\in\set{B}} \op{n}_{\ell\dn}
		\prod_{\ell\notin\set{B}} (1-\op{n}_{\ell\dn}) \ket{\psi_0^\dn}
\end{aligned}\end{equation}
\begin{equation}\begin{aligned} \label{eq: Jxy_raw}
\bra{\psi} \opdag{c}_{i\dn} \op{c}_{i\up} \opdag{c}_{j\up} \op{c}_{j\dn} \ket{\psi} 
	& = \nquad \sum_{\substack{ (\set{A},\set{B}) \\ i\in\set{A}, j\in\set{B}}} 
		y_{i\up} y_{j\up} y_{i\dn} y_{j\dn} 
		\left( \prod_{\ell\in\set{A}\setminus\{i\}} y_{\ell\up}^2 \right)
		\left( \prod_{\ell\in\set{B}\setminus\{j\}} y_{\ell\dn}^2 \right)
	  \\ & \quad \times
		\bra{\psi_0^\up} \op{c}_{i\up} \opdag{c}_{j\up}
		\nquad \prod_{\ell\in\set{A}\setminus\{i\}} \nquad \op{n}_{\ell\up}
		\!\! \prod_{\ell\notin\set{A}\cup\{j\}} \nquad (1-\op{n}_{\ell\up}) \ket{\psi_0^\up}
		\bra{\psi_0^\dn} \opdag{c}_{i\dn} \op{c}_{j\dn}
		\nquad \prod_{\ell\in\set{B}\setminus\{j\}} \nquad \op{n}_{\ell\dn}
		\!\! \prod_{\ell\notin\set{B}\cup\{i\}} \nquad (1-\op{n}_{\ell \dn}) \ket{\psi_0^\dn}
\end{aligned}\end{equation}
\end{widetext}
and which the results for $\bra{\psi} \opdag{c}_{j\dn} \op{c}_{i\dn} \ket{\psi}$, $\bra{\psi} \opdag{c}_{i\dn} \op{c}_{i\dn} \opdag{c}_{j\dn} \op{c}_{j\dn} \ket{\psi}$ and  $\bra{\psi} \opdag{c}_{i\up} \op{c}_{i\dn} \opdag{c}_{j\dn} \op{c}_{j\up} \ket{\psi}$ can be obtained from the above with the substitution $\up \leftrightarrow \dn$ and $\set{A} \leftrightarrow \set{B}$.

It is known\cite{Ogawa} that the expectations appearing in Eq.~\ref{eq: norm_raw}--\ref{eq: Jxy_raw} can be written in terms of determinants with elements of the form $\bra{\psi_0}\opdag{c}_i \op{c}_j\ket{\psi_0}$. To summarize,

\begin{align}
\label{eq: det norm} \bra{\psi_0^\sigma} \prod_{\ell\in\set{S}} \op{n}_{\ell\sigma}
		\prod_{\ell\notin\set{S}} (1-\op{n}_{\ell\sigma}) \ket{\psi_0^\sigma}
	& = \det U_{\set{S}\sigma} \\
\label{eq: det hop} \bra{\psi_0^\sigma} \opdag{c}_{j\sigma} \op{c}_{i\sigma}
		\nquad \prod_{\ell\in\set{S}\setminus\{i\}} \nquad \op{n}_{\ell\sigma}
		\prod_{\ell\notin\set{S}\cup\{j\}} \nquad (1-\op{n}_{\ell\sigma}) \ket{\psi_0^\sigma}
	& = \det U_{\set{S}\sigma}^{(ij)} 
\end{align}
where in Eq.~\ref{eq: det hop} it is assumed that $i\in\set{S}$ while $j\notin\set{S}$, and that in both equations it is assumed that the constraint $|\set{S}| = \sum_i \bra{\psi_0} \op{n}_{i\sigma} \ket{\psi_0}$ is met. $U_{\set{S}\sigma}$ is an $|\set{L}|$-by-$|\set{L}|$ matrix, whose entries are given by:
\begin{equation}
[U_{\set{S}\sigma}]_{ij} = \left\{ \begin{array}{ll}
\bra{\psi_0^\sigma} \opdag{c}_{i\sigma} \op{c}_{j\sigma} \ket{\psi_0^\sigma}  
	& \textrm{when }i\in\set{S}\\
\delta_{ij} - \bra{\psi_0^\sigma} \opdag{c}_{i\sigma} \op{c}_{j\sigma} \ket{\psi_0^\sigma}
  & \textrm{when }i\notin\set{S}
\end{array} \right.
\end{equation}
The matrix $U_{\set{S}\sigma}^{(ij)}$ is obtained from $U_{\set{S}\sigma}$ by first exchanging the i-th and j-th column of $U_{\set{S}\sigma}$ and then removing the j-th row and the j-th column from the resulting matrix.

\section{Gutzwiller Approximation and Thermodynamic Limit}

The precise content of the Gutzwiller approximation can now be stated. When evaluating $\bra{\psi} \op{Q} \ket{\psi}$, where $\op{Q}$ is expressed in terms of creation and annihilation operators only on sites $\set{I} = \{i_1,...,i_n\}$, we first expand the expectation in configuration basis as in Eq.~\ref{eq: norm_raw}--\ref{eq: Jxy_raw}. Then, each individual term in the expansion would depend on the details of the configuration. \emph{The Gutzwiller approximation amounts to neglecting the non-combinatorial dependence on sites other than $\{i_1,...,i_n\}$ in the quantum-mechanical part of each term. Hence it can be viewed as an approximation in evaluating the determinants $\det U_{\set{S}\sigma}$ and $\det U_{\set{S}\sigma}^{(ij)}$.} For a canonical wavefunction defined on a superlattice, by considering the permutation expansion of the determinants, it can be seen that in general the off-diagonal terms depend on the correlations in the configuration while the purely diagonal term depends only combinatorially on the configuration. In other words, if $\set{S}$ and $\set{S}'$ contain the same number of sites on each sublattice, then the purely diagonal term in the permutation expansion of $\det U_{\set{S}\sigma}$ and $\det U_{\set{S'}\sigma}$ agrees, while the off-diagonal terms in general disagree. The case for $\det U_{\set{S}\sigma}^{(ij)}$ is analogous. Hence, in the Gutzwiller approximation, we approximate:
\begin{equation} \begin{aligned}
\det U_{\set{S}\sigma} & \approx \det U_{\set{S}\sigma;\set{I}} \times
	\prod_{j\notin\set{I}} [U_{\set{S}\sigma}]_{jj}\\
\det U_{\set{S}\sigma}^{(ij)} & \approx \det U^{(ij)}_{\set{S}\sigma;\set{I}} \times 
	\prod_{j\notin\set{I}} [U_{\set{S}\sigma}]_{jj} 
\end{aligned} \end{equation}
where $U_{\set{S}\sigma;\set{I}}$ is the $n$-by-$n$ subblock of $U_{\set{S}\sigma}$ whose elements are those in $U_{\set{S}\sigma}$ that connects one lattice sites in $\set{I}$ to another, and that $U^{(ij)}_{\set{S}\sigma;\set{I}}$ is obtained from $U_{\set{S}\sigma;\set{I}}$ in the same way which $\det U_{\set{S}\sigma}^{(ij)}$ is obtained from $U_{\set{S}\sigma}$.

Applying the approximation, Eq.~\ref{eq: norm_raw}--\ref{eq: Jxy_raw} simplifies to:
\begin{widetext}
\begin{equation} \label{eq: norm_Gutz}
\avg{\psi|\psi} = \sum_{ (\set{A},\set{B}) } 
	\prod_{j\in\set{A}} y_{j\up}^2 n_{j \up}
	\prod_{j\notin\set{A}} (1-n_{j \up})
	\prod_{j\in\set{B}} y_{j\dn}^2 n_{j \dn}
	\prod_{j\notin\set{B}} (1-n_{j \dn})
\end{equation}
\begin{equation}\label{eq: density_Gutz}
\bra{\psi} \opdag{c}_{i\up} \op{c}_{i\up} \ket{\psi} 
	= \sum_{\substack{(\set{A},\set{B}) \\ i\in \set{A}}} 
	\prod_{j\in\set{A}} y_{j\up}^2 n_{j \up}
	\prod_{j\notin\set{A}} (1-n_{j \up})
	\prod_{j\in\set{B}} y_{j\dn}^2 n_{j \dn}
	\prod_{j\notin\set{B}} (1-n_{j \dn})
\end{equation}
\begin{equation} \begin{aligned} \label{eq: hop_Gutz}
\bra{\psi} \opdag{c}_{j\up} \op{c}_{i\up} \ket{\psi} 
	& =\nquad \sum_{\substack{ (\set{A},\set{B}) \\ i\in \set{A}, j\notin \set{A}\cup\set{B}}} 
		\nquad y_{i\up} y_{j\up} \avg{\opdag{c}_{j\up} \op{c}_{i\up}}_0
		\prod_{\ell\in\set{A}\setminus\{i\}} \nquad y_{\ell\up}^2 n_{\ell\up} \nquad
		\prod_{\ell\notin\set{A}\cup\{j\}} \nquad (1-n_{\ell\up})
		\prod_{\ell\in\set{B}} y_{\ell\dn}^2 n_{\ell\dn}
		\prod_{\ell\notin\set{B}} (1-n_{\ell\dn})
\end{aligned}\end{equation}
\begin{equation}\begin{aligned} \label{eq: Jz1_Gutz}
\bra{\psi} \opdag{c}_{i\up} \op{c}_{i\up} & \opdag{c}_{j\up} \op{c}_{j\up} \ket{\psi} 
	= \sum_{\substack{(\set{A},\set{B}) \\ i,j\in \set{A}}}
		y_{i\up}^2 y_{j\up}^2 \avg{\opdag{c}_{i\up} \op{c}_{i\up} \opdag{c}_{j\up} \op{c}_{j\up}}_0
		\prod_{\ell\in\set{A}\setminus\{i,j\}} \nquad y_{\ell\up}^2 n_{\ell\up}
		\prod_{\ell\notin\set{A}} (1-n_{\ell\up})
		\prod_{\ell\in\set{B}} y_{\ell\dn}^2 n_{\ell\dn}
		\prod_{\ell\notin\set{B}} (1-n_{\ell\dn})
\end{aligned}\end{equation}
\begin{equation}\begin{aligned} \label{eq: Jz2_Gutz}
\bra{\psi} \opdag{c}_{i\up} \op{c}_{i\up} & \opdag{c}_{j\dn} \op{c}_{j\dn} \ket{\psi} 
	= \sum_{\substack{(\set{A},\set{B}) \\ i\in \set{A}, j\in \set{B}}}
		y_{i\up}^2 y_{j\dn}^2 \avg{\opdag{c}_{i\up} \op{c}_{i\up} \opdag{c}_{j\dn} \op{c}_{j\dn}}_0
		\prod_{\ell\in\set{A}\setminus\{i\}} \nquad y_{\ell\up}^2 n_{\ell\up}
		\prod_{\ell\notin\set{A}} (1-n_{\ell\up}) \nquad
		\prod_{\ell\in\set{B}\setminus\{j\}} \nquad y_{\ell\dn}^2 n_{\ell\dn}
		\prod_{\ell\notin\set{B}} (1-n_{\ell\dn})
\end{aligned}\end{equation}
\begin{equation}\begin{aligned} \label{eq: Jxy_Gutz}
\bra{\psi} \opdag{c}_{i\dn} \op{c}_{i\up} & \opdag{c}_{j\up} \op{c}_{j\dn} \ket{\psi} 
	= \nquad \sum_{\substack{ (\set{A},\set{B}) \\ i\in\set{A}, j\in\set{B}}} 
		y_{i\up} y_{j\up} y_{i\dn} y_{j\dn} 
		\avg{\opdag{c}_{i\dn} \op{c}_{i\up} \opdag{c}_{j\up} \op{c}_{j\dn}}_0
		\prod_{\ell\in\set{A}\setminus\{i\}} \nquad y_{\ell\up}^2 n_{\ell\up}
		\nquad \prod_{\ell\notin\set{A}\cup\{j\}} \nquad (1-n_{\ell\up})
		\nquad \prod_{\ell\in\set{B}\setminus\{j\}} \nquad y_{\ell\dn}^2 n_{\ell\dn}
		\nquad \prod_{\ell\notin\set{B}\cup\{i\}} \nquad (1-n_{\ell\dn})
\end{aligned} \end{equation}
where $n_{j\sigma} = \avg{\op{n}_{j\sigma}}_0$ in above and henceforth.

With our assumption of a superlattice structure, the lattice can be divided into $\kappa$ sublattices. Let $M$ = $|\set{L}|/\kappa$ be the number of unit supercell, then a little bit of combinatorics (c.f.\@ Eq.~\ref{eq: classical weight} and \ref{eq: config count}) gives:
\begin{equation} \begin{aligned}
\label{eq: norm_Gutz2}
\avg{\psi|\psi} & = 
	\sum_{\{a_{I\sigma}\}} \prod_{I} \bigg(
	\frac{M!}{a_{I\up}! \, a_{I\dn}! \, (M-a_{I\up}-a_{I\dn})!} \bigg)
	y_{I\up}^{2 a_{I\up}} n_{I\up}^{a_{I\up}} (1-n_{I\up})^{M-a_{I\up}}
	\, y_{I\dn}^{2 a_{I\dn}} n_{I\dn}^{a_{I\dn}} (1-n_{I\dn})^{M-a_{I\dn}}
	\\ & 
	= \sum_{\{a_{I\sigma}\}} F(\{a_{I\sigma}\})
\end{aligned}\end{equation}
\end{widetext}
where $I$ label the sublattices, and $F(\{a_{I\sigma}\})$ is defined in the obvious way. The sum is over all non-negative sets of $\{a_{I\up}\}$ and $\{a_{I\dn}\}$ such that $\sum_I a_{I\up} = N_{\up}$ and $\sum_I a_{I\dn} = N_{\dn}$. In a similar manner, it can be checked that Eq.~\ref{eq: density_Gutz}--\ref{eq: Jxy_Gutz} can be written in the general form $\sum_{\{a_{I\sigma}\}} (\ldots) F(\{a_{I\sigma}\})$. Specifically, for site-$i$ is on sublattice $P$ and site-$j$ is on sublattice $Q$, with $P$ and $Q$ distinct,
\begin{equation}
\label{eq: density_Gutz2}
\bra{\psi} \op{n}_{i\sigma} \ket{\psi} =
	\! \sum_{\{a_{I\sigma}\}} \! \frac{a_{P\sigma}}{M} F(\{a_{I\sigma}\})
\end{equation}
\begin{equation}\begin{aligned}
\label{eq: hop_Gutz2}
\bra{\psi} & \opdag{c}_{j\sigma} \op{c}_{i\sigma} \ket{\psi} = 
	\avg{\opdag{c}_{j\sigma} \op{c}_{i\sigma}}_0 \\
	& \times \sum_{\{a_{I\sigma}\}} \! \frac{a_{P\sigma}(M-a_{Q\up}-a_{Q\dn})y_{P\sigma}y_{Q\sigma}}
	{M^2 (y_{P\sigma}^2 n_{P\sigma}) (1-n_{Q\sigma})}
	F(\{a_{I\sigma}\})
\end{aligned}\end{equation}
\begin{equation}\begin{aligned}
\label{eq: Jxy_Gutz2}
\bra{\psi} & \opdag{c}_{i\up}\op{c}_{i\dn} \opdag{c}_{j\dn}\op{c}_{j\up} \ket{\psi} =	\avg{\opdag{c}_{i\up}\op{c}_{i\dn} \opdag{c}_{j\dn}\op{c}_{j\up}}_0 \\
	& \times \sum_{\{a_{I\sigma}\}} \!
	\frac{a_{P\dn} a_{Q\up} y_{P\up} y_{P\dn} y_{Q\up} y_{Q\dn} F(\{a_{I\sigma}\})}
	{M^2 (1-n_{P\up}) (y_{P\dn}^2 n_{P\dn}) (1-n_{Q\dn}) (y_{Q\up}^2 n_{Q\up})}
\end{aligned}\end{equation}
\begin{equation}
\label{eq: Jz1_Gutz2}
\bra{\psi} \opdag{c}_{i\sigma}\op{c}_{i\sigma} \opdag{c}_{j\sigma} \op{c}_{j\sigma} \ket{\psi} =
	\avg{\opdag{c}_{i\sigma}\op{c}_{i\sigma} \opdag{c}_{j\sigma} \op{c}_{j\sigma}}_0
	\! \sum_{\{a_{I\sigma}\}} \! \frac{a_{P\sigma} a_{Q\sigma} F(\{a_{I\sigma}\})}
	{M^2 n_{P\sigma} n_{Q\sigma}}
\end{equation}
\begin{equation}
\label{eq: Jz2_Gutz2}
\bra{\psi} \opdag{c}_{i\sigma}\op{c}_{i\sigma} \opdag{c}_{j\nn{\sigma}} \op{c}_{j\nn{\sigma}} \ket{\psi} =
	\! \sum_{\{a_{I\sigma}\}} \! \frac{a_{P\sigma} a_{Q\nn{\sigma}}}{M^2}
	F(\{a_{I\sigma}\})
\end{equation}
The corresponding expressions when $P$ and $Q$ are identical are similar, and yield the same formula once the thermodynamic limit is taken (Eq.~\ref{eq: density_thermal}--\ref{eq: Jxy_thermal} below).

As discussed in the main text, in the thermodynamic limit, whereby $|\set{L}| \rightarrow \infty$, the sum $\sum_{\{a_{I\sigma}\}}$ can be replaced by the single term in which $F(\{a_{I\sigma}\})$ attains maximum under the constraints $\sum_I a_{I\sigma} = N_{\sigma}$. The maximization condition results in Eq.~\ref{eq: rho} in the main text, with $\rho_{I\sigma}$ formally defined as $\rho_{I\sigma} = a_{I\sigma}/M$ in the present context. Substituting the $\rho_{i\sigma}$ into Eq.~\ref{eq: norm_Gutz2}--\ref{eq: Jz2_Gutz2} then gives:
\begin{align}
\label{eq: density_thermal}
\avg{\op{n}_{i\sigma}} & = \rho_{P\sigma} \\
\label{eq: exchange_thermal}
\avg{\opdag{c}_{j\sigma} \op{c}_{i\sigma}} & = 
	\frac{y_{Q_\sigma}}{y_{P_\sigma}}
	\frac{\rho_{P\sigma} (1-\rho_{Q\up}-\rho_{Q\dn})}{n_{P\sigma} (1-n_{Q\sigma}) }
	\avg{\opdag{c}_{j\sigma} \op{c}_{i\sigma}}_0 \\
\label{eq: Jz_thermal}
\avg{\opdag{c}_{i\sigma} \op{c}_{i\sigma} \opdag{c}_{j\sigma'} \op{c}_{j\sigma'}} & =
	\frac{\rho_{P\sigma} \, \rho_{Q\sigma'}}{n_{P\sigma} n_{Q\sigma'}}
	\avg{\opdag{c}_{i\sigma} \op{c}_{i\sigma} \opdag{c}_{j\sigma'} \op{c}_{j\sigma'}}_0 \\
	\label{eq: Jxy_thermal}
\avg{\opdag{c}_{i\sigma} \op{c}_{i\nn{\sigma}} \opdag{c}_{j\nn{\sigma}} \op{c}_{j\sigma}} & =
	\frac{y_{P\sigma} y_{Q\nn{\sigma}}}{y_{P\nn{\sigma}} y_{Q\sigma}}
	\frac{\rho_{P\nn{\sigma}} \, \rho_{Q\sigma} 
		\avg{\opdag{c}_{i\sigma} \op{c}_{i\nn{\sigma}} \opdag{c}_{j\nn{\sigma}} \op{c}_{j\sigma}}_0}
	{n_{P\nn{\sigma}} (1-n_{P\sigma}) n_{Q\sigma} (1-n_{Q\nn{\sigma}})}
\end{align}
where site-$i$ is on sublattice $P$ while site-$j$ is on sublattice $Q$ ($P$ and $Q$ may be identical). Using Eq.~\ref{eq: rho}, it can be checked explicitly that the hermiticity of expectation values is preserved under the approximation. Specifically, the R.H.S. of Eq.~\ref{eq: exchange_thermal} and Eq.~\ref{eq: Jxy_thermal} remain unchanged upon $P \leftrightarrow Q$. This allows us to eliminate the the $y_{I\sigma}$ in these equations. Upon simplifications and rearrangements this yields Eq.~\ref{eq: hop}--\ref{eq: Jdn} in the main text.

\end{document}